\documentclass[pre,onecolumn,superscriptaddress,aps,notitlepage,10pt]{revtex4-1}

\usepackage{mathtools}
\usepackage{newpxtext,newpxmath}
\usepackage[utf8]{inputenc}
\usepackage{hyperref}
\usepackage{graphicx}
\usepackage{lipsum}
\usepackage[normalem]{ulem}
\usepackage[margin=1.25in]{geometry}
\usepackage{subcaption}
\usepackage[dvipsnames]{xcolor}
\usepackage{quantikz}
\usetikzlibrary{shapes}
\usepackage{algorithm}
\usepackage{color}
\usepackage{algpseudocode}
\usepackage{comment}
\usepackage{enumitem}
\usepackage{url}
\usepackage{float}
\usepackage[textwidth=2.5cm]{todonotes}
\usepackage{adjustbox}
\usepackage{multirow}

\setuptodonotes{fancyline, color=blue!30}
\usepackage{outlines}

\begin{document}
\begin{abstract}
     Quantum computing promises revolutionary advances in modeling materials and molecules. However, the up-to-date runtime estimates for utility-scale applications on certain quantum hardware systems are in the order of years rendering quantum computations impractical. Our work incorporates state-of-the-art innovations in all key aspects of the fault-tolerant quantum computing (FTQC) stack to show how quantum computers could realistically and practically tackle CO$_2$ utilization for green energy production. We bring down the quantum computation runtime from $22$ years to just $1$ day, achieving a significant $7.9e03$ reduction from previous state-of-the-art work. This reduction renders the quantum computation feasible, challenges state-of-the-art classical methods and results to a predicted run-time quantum advantage. We provide a rigorous analysis of how different innovations across the stack combine to provide such reductions. Our work provides strong evidence that all layers of FTQC are crucial in the quest for quantum advantage. Our analysis can be applied to related problems on FTQC and for any type of quantum architecture. Our methodology connects quantum algorithms to applications of positive real-world impact and leads to compelling evidence of achievable quantum advantage. 
\end{abstract}

\title{Achieving Utility-Scale Applications through Full Stack Co-Design of \newline Fault Tolerant Quantum Computers}

\author{Katerina Gratsea}
\email{gratsea.katerina@gmail.com}
\affiliation{University of Wisconsin -- Madison}

\author{Matthew Otten}
\affiliation{University of Wisconsin -- Madison}

\maketitle

\tableofcontents

\section{Introduction}

One of the most promising applications of quantum computing lies in material and molecular simulations~\cite{goings2022reliably}. Specifically, quantum ground state energy estimation (GSEE) algorithms are extensively studied in the field~\cite{DFTHC} and hold the potential to provide highly accurate estimations with significant industrial impact~\cite{DF_pyLIQTR, goings2022reliably, bellonzi2024}. To date, most works have focused on comparing quantum and classical algorithms primarily through complexity-theoretic arguments~\cite{DFTHC, DF_pyLIQTR}, while only just a few works have attempted to provide an estimate of the runtime and physical qubit counts of the quantum computation by incorporating a full Quantum Resource Estimation (QRE) analysis~\cite{beverland2022, bellonzi2024, scaling, otten2024qrechemquantumresourceestimation}.

As fault-tolerant quantum computing continues to advance, it is becoming increasingly clear that distributed quantum architecture modeling will be essential for realizing the millions of physical qubits required to support industrially relevant computations~\cite{scaling, Simon}. To achieve more accurate and realistic resource estimates, the distributed architecture model must be incorporated into QRE analyses. Equally important, however, is the integration of all other critical layers of fault-tolerant quantum computing—namely the algorithmic layer, the logical processing layer, and quantum error correction (QEC)—since each of these has a substantial impact on both overall runtime and physical qubit requirements.

In contrast to prior work~\cite{beverland2022, otten2024qrechemquantumresourceestimation}, which employed simplified hardware modeling across multiple quantum platforms to estimate runtimes and physical qubit counts, our approach focuses on a specific quantum hardware system. Notably, we select the system that had the poorest performance in terms of runtime and demonstrate that by incorporating advances across all layers of the computational stack in parallel can lead to substantial reductions in overall runtime estimates. Through a more accurate and realistic Quantum Resource Estimation analysis, we underscore the critical role of full-stack co-design in enabling fault-tolerant quantum computing for utility-scale applications.

Towards that end, we integrate three open source software tools, namely pyLIQTR~\cite{Obenland_pyLIQTR}, Bench-Q~\cite{benchq} and pyZX~\cite{kissinger2020Pyzx}. Together, these enable a comprehensive workflow: applications are first mapped to quantum circuits using pyLIQTR, further optimized through ZX-calculus with pyZX, and then compiled with graph-state methods while incorporating QEC schemes and detailed distributed hardware modeling via Bench-Q.  Importantly, we not only utilized but also contributed to the development of these open-source tools and the integration of their distinct functionalities into a unified pipeline. 

Our methodology represents an end-to-end QRE workflow, incorporating both detail and scale. Our analysis extends well beyond the system sizes studied in earlier studies: in particular, we analyze large catalyst systems of industrial relevance with up to 150 orbitals and benchmark our analysis against state-of-the-art classical simulations that we performed using the SHCI method~\cite{Li2018}. This approach allows us to draw robust conclusions about the emergence of quantum advantage.

The structure of the work is as follows. In Sec.~\ref{Sec:Two}, we define the application as a computer science problem relevant for any computational method, classical or quantum, along with the relevance of the studied problem for a positive real-world impact. In Sec.~\ref{Sec:Novelty}, we discuss the main methodology used and provide a summary on the results of the work. In the section following, we discuss in more detail the observed quantum advantages of the quantum computation over classical, namely on the accuracy (Sec.~\ref{Subsec:Accuracy}) and runtime (Sec.~\ref{subsec:runtime}). In Sec.~\ref{Sec:Classical_Benchmarking}, we discuss in detail the classical benchmarking of the studied problems along with the relevant bottlenecks, while in Sec.~\ref{Sec:Viability} we discuss the viability of the quantum solution. Finally, we present the conclusions and outlook in Sec.~\ref{Sec:Conclusions}.

\section{Problem Definition} \label{Sec:Two}

\subsection{Definition}\label{Sec:Problem_statement}

Eigenstate energy estimation of systems taking part in complex chemical reactions is a crucial step for many industries in pharmaceutical, material, and climate change. A reaction that has extensively been studied with classical chemistry methods is the transformation of carbon dioxide, as it is a greenhouse gas that is a major contributor to climate change~\cite{01_Review_CO2, 02_Review_CO2}. Finding ways to reduce carbon dioxide will have significant positive societal impacts benefiting more than one of the United Nations Sustainable Development Goals~\cite{UNSDG2030}. 

Limiting or even inverting rising carbon dioxide levels in the atmosphere is an important goal and all possible ways to accomplish it must be considered~\cite{Hepburn2019CO2}. One approach involves chemically transforming carbon dioxide through chemical reactions~\cite{complex_XVIII}. Ideally, a very favorable pathway would transform CO$_2$ with green H$_2$ to renewable fuels and raw materials ~\cite{sustainable_conversion} as suggested below:

\begin{equation}
\mathrm{CO_2} \;+\; \mathrm{H_2}
\xrightarrow{\text{catalyst}}\;
\{\text{renewable fuels}\} \;+\; \{\text{raw materials}\}.
\label{eq:generic_CO2_H2}
\end{equation}

Identifying such a pathway is very difficult. Currently, different proposed pathways face several challenges, such as a mix of products being produced, catalysts being deactivated or catalysts being produced on raw materials. A promising pathway that has been identified is the following:

\begin{equation}
\mathrm{CO_2} \;+\; \mathrm{H_2}
\xrightarrow{\text{Ru-based}}\;
\mathrm{CH_3OH} \;+\; \mathrm{H_2O}.
\label{eq:methanol_CO2_H2}
\end{equation}

\begin{figure}[h!]
    \centering
    \includegraphics[width=1.0\textwidth]{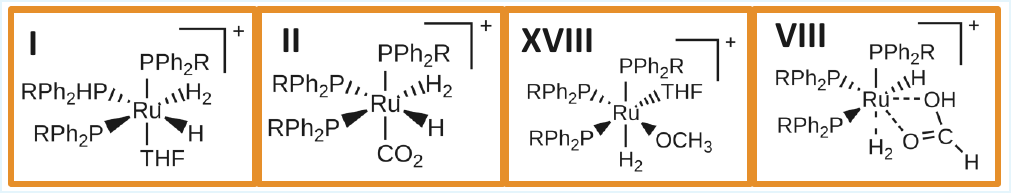}
    \caption{Examples of Ru-based catalyst structures that are relevant in the carbon capture catalytic cycle.}
    \label{Fig.1-Catalysts}
\end{figure}

This pathway is seen as extremely favorably~\cite{sustainable_conversion} as it involves hydrogen (H$_2$) which could potentially come from renewable sources (e.g. from water electrolysis powered by solar/wind), thus making the whole process carbon-neutral~\cite{Saez2022}. Moreover, methanol (CH$_3$OH) is considered a renewable fuel that could directly be used in combustion engines or power generation~\cite{methanol_economoy, Saez2022}. This plays a key role in carbon recycling: CO$_2$ is captured, converted, used, and then re-emitted — but no new fossil carbon is introduced. This is referred to as a closed carbon loop. While the benefits of the pathway are obvious, the reaction mechanism in Eq.~\eqref{eq:methanol_CO2_H2} only suggests that the catalyst is Ru-based, but in practice the reaction involves a series of reactions with different Ru-based catalyst structures (see Fig.~\ref{Fig.1-Catalysts}).

In more detail, the formation of methanol (CH$_3$OH) and water occurs through hydrogenation and ligand exchanges with transition states (i.e. II and VIII shown in Fig.~\ref{Fig.1-Catalysts}) marking energy barriers between key species. Starting from complex I, the cycle involves sequential reactions with H$_2$, CO$_2$, and water, forming intermediates II through XVIII. In Fig.~\ref{Fig.1-Catalysts} we show some of the intermediate species involved (see Ref.~\cite{complex_XVIII} for a complete list). The necessity for intermediate species arises from the need to incrementally break and form bonds (e.g., C=O to C-H) in a thermodynamically and kinetically feasible manner. One intermediate reaction step in the catalytic pathway is

\begin{equation}\label{Eq:Reaction}
\mathrm{I} + \mathrm{CO_2} + 2\,\mathrm{H_2} \rightarrow \mathrm{XVIII} + \mathrm{H_2O}.
\end{equation}
Both I and XVIII are Ru complexes, so the catalysts are reacting species themselves. This is why ``Ru-based" is not written above the arrow. In this reaction step (Eq.~\eqref{Eq:Reaction}), complex I reacts with  CO$_2$ and two H\textsubscript{2} to form complex XVIII and water, with the methanol moiety remaining coordinated to the Ru center of complex XVIII as an --OCH\textsubscript{3} ligand. 

The intermediate reaction step shown in Eq.~\eqref{Eq:Reaction} is between two key catalytic intermediates: complex I and complex XVIII. The relative electronic energy between those two key catalytic intermediates is
\begin{equation}\label{Eq:DE}
     \Delta E^{\text{el}}_{\text{rel}}= E^{\text{el}}(\text{XVIII}) + E^{\text{el}}(\text{H}_2\text{O}) - 2 \cdot E^{\text{el}}(\text{H}_2) - E^{\text{el}}(\text{I}) - E^{\text{el}}(\text{CO}_2).
\end{equation}
Eq.~\eqref{Eq:DE} captures the energy change in a sub-reaction within the catalytic cycle, not the full release of CH\textsubscript{3}OH. It serves as a computational proxy for benchmarking quantum algorithms or classical methods on a chemically meaningful but reduced portion of the full catalytic process. Additional calculations of energy differences are necessary to fully optimize the catalytic cycle. Ultimately, these energy differences inform calculations of the catalytic rates, which determine the overall efficiency of the cycle.

Thus, understanding the catalytic cycles requires high-accuracy energy calculations of the catalysts and all other systems participating in the reaction. Throughout this work, we study the Ru-based catalyst referred to as compound XVIII~\cite{complex_XVIII, beverland2022}, which is one of the largest and most computationally difficult candidates involved in the chemical transformation of Eq.~\eqref{eq:methanol_CO2_H2}. 

Given the Hamiltonian that accurately represents the compound XVIII the concrete problem is to estimate the ground state energy to sufficient accuracy (which we take to be 1mHa). Providing a solution to this problem with any method (classical or quantum) can unlock the study of the chemical reactions around carbon dioxide and elucidate mechanisms for limiting rising carbon dioxide levels in the atmosphere.

\subsection{Impact} \label{Sec:Impact}

The carbon dioxide utilization by chemical transformation necessitates energy estimations within a small enough accuracy (which we take to be 1 mHa) to elucidate mechanisms for limiting rising carbon dioxide levels in the atmosphere. Energy estimations have been studied extensively with different classical methods both computationally and experimentally for different Ru-based catalysts~\cite{complex_XVIII}. The strong electron-electron correlations that govern the Ru-based catalysts create bottlenecks for classical methods. Namely, the electronic energies derived by distinct classical methods differ by far more than the widely accepted chemical accuracy of 1.6mHa~\cite{complex_XVIII}. 

If we had solutions to the presented bottlenecks of the classical computation, it would mean that the catalyst discovery in silico would be possible. This would have profound implications in understanding the mechanism that governs the efficient transformation of carbon dioxide. Converting CO$_2$ to methanol represents one of the most promising routes and is already industrially demonstrated with pilot and commercial plants operational worldwide for Cu/ZnO-based catalysts~\cite{Yang2022} and accurately estimating the ground state energies of Ru-based catalysts can help unlock the ruthenium-based CO$_2$ hydrogenation to methanol.

\begin{figure}[h!]
    \centering
    \includegraphics[width=1.1\textwidth]{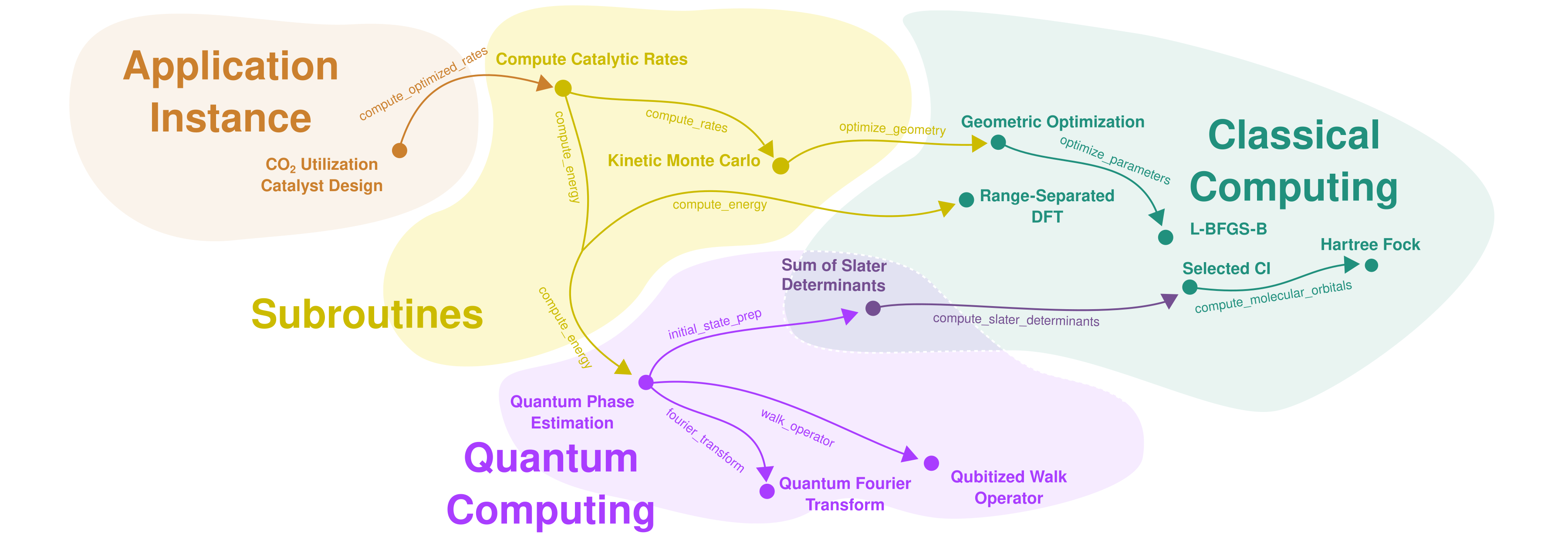}
    \caption{Quantum Benchmarking Graph (QBG) of the proposed computational workflow. }
    \label{Fig.QBG}
\end{figure}

In Fig.~\ref{Fig.QBG}, the workflow is expressed in more detail using the Quantum Benchmarking Graph (QBG) formalism; a structured, graphical technique designed to systematically break down an application instance into its core subroutines and fundamental computational tasks~\cite{Otten1, Otten2}. A QBG takes the form of an attributed, directed acyclic graph, where both nodes and edges convey computational roles and data dependencies. As it is shown in Fig.~\ref{Fig.QBG}, to compute the catalytic reaction rates, we first need to compute the electronic energies of the reactants and this is where quantum computing is needed for the most challenging systems. Then, additional free energy calculations are necessary, where range-separated DFT can be applied~\cite{complex_XVIII}. The full workflow is shown diagrammatically in Fig.~\ref{Fig.QBG}.

\begin{figure}[h!]
    \centering
    \includegraphics[width=1.0\textwidth]{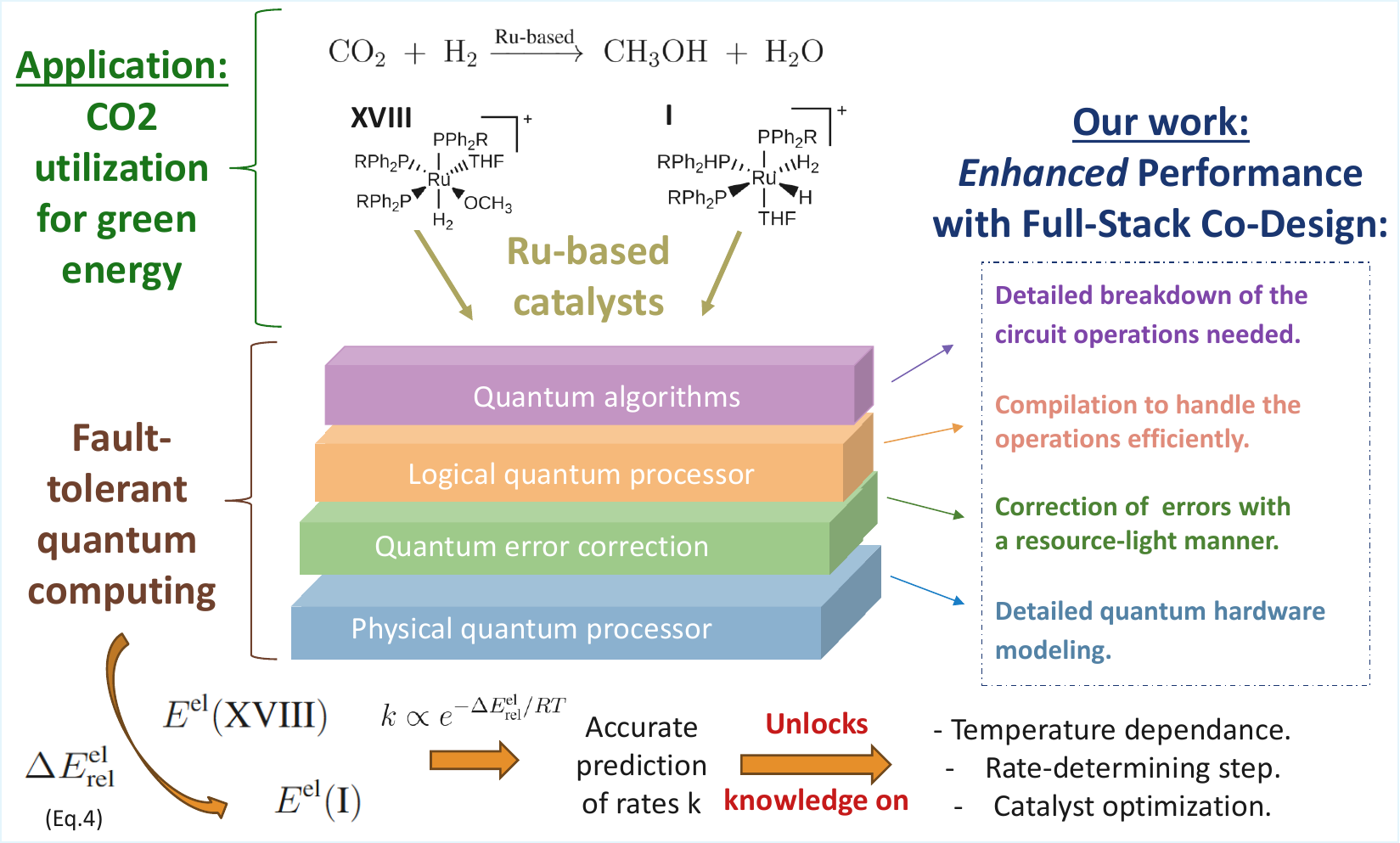}
    \caption{The computation of accurate electronic energies on fault-tolerant quantum computers (FTQC) will help to accurately predict chemical reaction rates. This will unlock the understanding on reaction mechanisms that determine the catalytic process and allow effective CO$_2$ utilization for green energy production. FTQC is an issue of great complexity with different layers involved and many choices available on all these layers of computation. Our full-stack co-design provides a rigorous analysis on the interplay of improvements across the stack for an achievable quantum advantage. }
    \label{Fig.2-FTQC_complexity}
\end{figure}

As already emphasized, the energy estimations govern the accurate prediction of reaction rates $k$,
\begin{equation}\label{Eq:rate_k}
    k \propto e^{-\Delta E^{\text{el}}_{\text{rel}}/R T},
\end{equation}
where $\Delta E^{\text{el}}_{\text{rel}}$ is given by Eq.~\eqref{Eq:DE} and $T$ refers to the temperature of the reaction. The exponential dependence of $\Delta E^{\text{el}}_{\text{rel}}$ demands the calculation of the energy differences to be sufficiently accurate; otherwise, even qualitative predictions are incorrect. Accurately predicting the catalytic reaction rates $k$ will, for example, unlock the understanding of temperature dependence of the reaction along with other important mechanisms, such as the rate-determining step that determines the speed of the reaction (see Fig.~\ref{Fig.2-FTQC_complexity}). Moreover, it will enable geometric structure optimization of the catalysts to potentially yield an even more favorable performance, such as long-term stability (see Fig.~\ref{Fig.2-FTQC_complexity}).

The knowledge gained would represent a major leap forward in carbon utilization technologies. Unlike traditional Cu/ZnO catalysts that require high pressures and temperatures, Ru-based systems--particularly homogeneous catalysts--can operate under milder and more energy-efficient conditions, potentially reducing the overall energy input for methanol production. If these catalysts could be engineered for high turnover frequency, long-term stability, and compatibility with green hydrogen, they would enable modular methanol production directly from CO$_2$ and renewable H$_2$. Therefore, the pathway of ruthenium-based CO$_2$ hydrogenation to methanol powered by renewable hydrogen has the potential to close the carbon loop, displacing fossil-derived methanol and contributing directly to climate goals~\cite{Stangeland2017}. The broader impact would be a significant acceleration of the transition to a carbon-neutral chemical economy.

Quantum algorithms for ground state energy estimation~\cite{Katabarwa_2024} of catalyst systems have attracted significant attention~\cite{complex_XVIII, bellonzi2024, beverland2022, DFTHC}. This works introduces a methodology that systematically analyzes the quantum algorithm performance and leads to observed evidence of quantum advantage. We incorporate the fault-tolerant complexity (see Fig.~\ref{Fig.2-FTQC_complexity}) in all layers (algorithmic, logical quantum processor, quantum error correction and physical quantum processor) with realistic assumptions to get accurate quantum resource estimates on the number of physical qubits and runtime that will be needed to run the quantum computation. Our work makes the compelling case that the proposed FTQC workflow can provide quantum advantage over the current state-of-the-art classical methods.

\section{Results} \label{Sec:Novelty}

Our work provides a methodology for understanding the full-stack resources required for a quantum computer to reach quantum advantage for the already established application on carbon dioxide utilization by chemical transformation.
As already discussed in detail in the above two sections, the quantum computer will take on the task of providing accurate ground state energy estimations (GSEE)~\cite{complex_XVIII, DFTHC}. 

Previous Quantum Resource Estimation (QRE) analysis suggested that 130 years~\cite{beverland2022} will be needed for the computation to run on ion-trap quantum hardware, which is not acceptable for any computational model. This estimation used a surface code cycle (SCC) of 6e-4 seconds and 1e-4 qubit gate error rate. Even if we use the up-to-date realistic (1e-4 seconds) and optimistic (1e-5 seconds) SCC times~\cite{Simon}, the total runtime of the quantum computation is reduced to 22 years and 2.2 years, respectively. Thus, even for the more optimistic SCC time, the total runtime is still rendering the attempt of getting accurate energy estimations on ion-trap quantum hardware unrealistic. Even worse, these simplistic estimates of 22 and 2.2 years exclude the increased number of communication ions that are needed in a detailed quantum hardware modeling of a modular architecture to support these faster runtimes.

In this work, we put forward a methodology in the service of connecting applications to quantum algorithms to quantum hardware by providing accurate and realistic QRE analysis that incorporate all aspects of fault-tolerant quantum computation (algorithmic, logical processor, QEC and physical processor) (see Sec.~\ref{Sec:Viability}, Sec.~\ref{Sec:Quantum_Advantage} and App.~\ref{Sec:Evidence_quantum_advantage}). 

\begin{figure}[h!]
    \centering
    \includegraphics[width=1.0\textwidth]{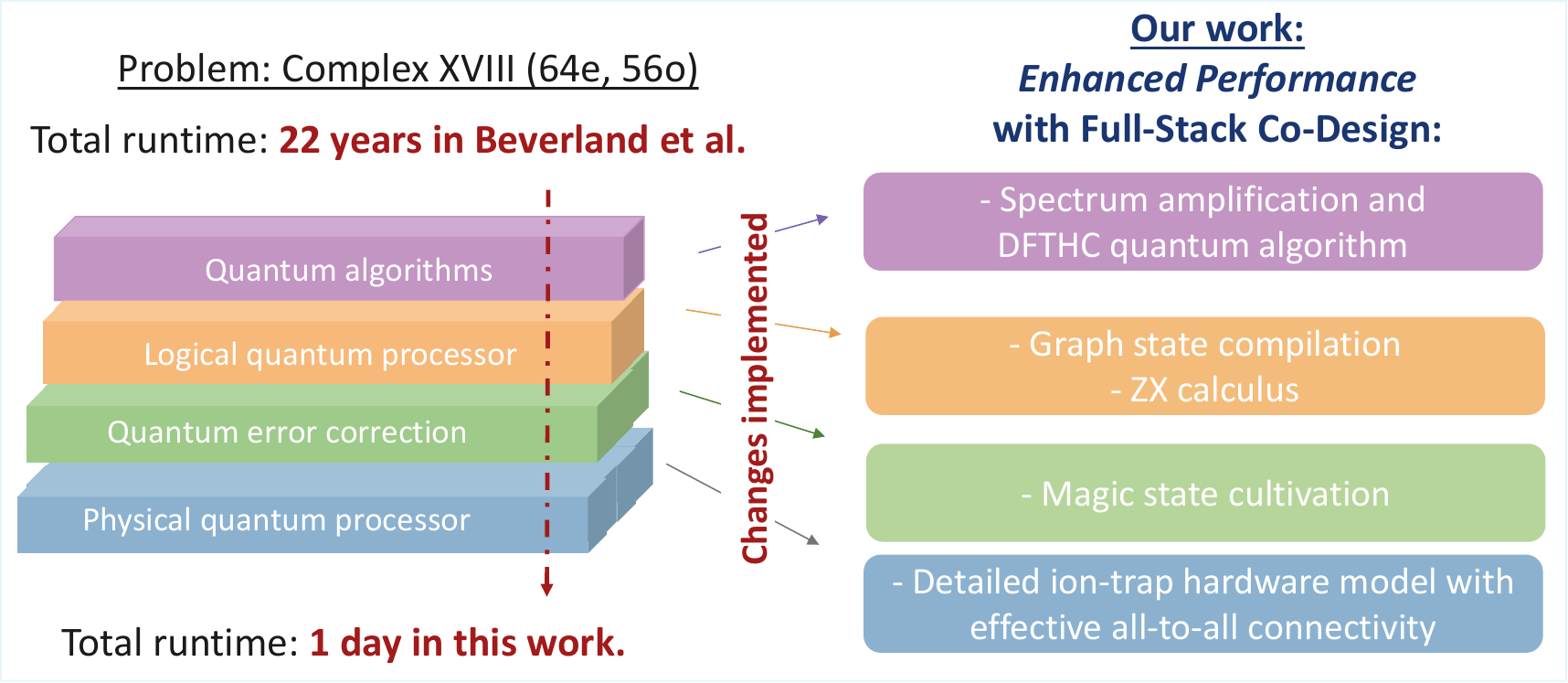}
    \caption{This figure shows in a nutshell all the advancements from different layers of the FTQC incorporated in this work that were necessary to reduce the runtime estimate by $7.3e03$. }
    \label{Fig.4-FTQC_advancements_detailed}
\end{figure}

We use proxy QASM quantum circuits of a utility-scale system and incorporates the current state-of-the art innovations in all key aspects of FTQC stack (see Fig.~\ref{Fig.4-FTQC_advancements_detailed}) to show how quantum computers could realistically tackle a problem of significant positive impact in our society. Importantly, we bring down the quantum computation runtime from 22 years to just 1 day, achieving a significant 7.9e03 improvement from previous state-of-the-art work~\cite{beverland2022}. Our analysis incorporates a detailed quantum hardware modeling of a modular, distributed architecture that realistically supports the 1e-4 second SCC time. Thus, more strictly, we find a reduction of 4.7e04 when comparing the basic modeling with 130 years runtime and a 6e-4 second SCC time reported in the work of Beverland et al.~\cite{beverland2022} to our analysis, which results in a runtime of just 1 day.

To accurately capture the complexity of FTQC in all the layers of computation involved, we incorporated different open-source software packages (pyLIQTR~\cite{Obenland_pyLIQTR}, pyZX~\cite{kissinger2020Pyzx} and Bench-Q~\cite{benchq}). Importantly, this work can be applied beyond the studied problem as the methodology used to reduce the runtime estimates could be applied to a broader area of problems even beyond GSEE and to more quantum hardware models. To demonstrate this, we also implemented a neutral-atom hardware model~\cite{sunami2025transversal} with 6e-4 second SCC time and 1e-4 qubit gate error rate (see Table~\ref{Table I}).

In our analysis, a handful of choices had to be made on the innovations to incorporate the ones that would have the most impact in reducing the runtime of the quantum computation, while still being based on realistic assumptions of how they could be realized. The choice on the innovations was a continuous feedback loop process on performing the QRE analysis, understanding the overheads of quantum computation and exploring potential ways that could be included to realistically reduce the runtime. Moreover, our work has a careful incorporation of a detailed quantum hardware modeling along with advances in all other layers of quantum computation listed in Fig.~\ref{Fig.4-FTQC_advancements_detailed}.

Moreover, to understand the limits of quantum advantage, we performed classical simulations with the selected heat-bath configuration interaction (SHCI) method (see Sec.~\ref{Sec:Classical_Benchmarking}) and provide detailed, state-of-the-art classical energy estimations for the Ru-based catalysts. In this section, we focused on the $\{56o, 64e\}$ system size for an easier comparison with the previous state-of-the-art work~\cite{beverland2022}, but in the next sections we extend the analysis to $\{100o, 100e\}$ and $\{150o, 150e\}$ ~---~pushing beyond what was previously done in either quantum or classical simulations for the studied system~\cite{complex_XVIII}.

\begin{table}[h!]
    \centering
    \vspace{6pt}
    \begin{tabular}{|l|l|l|l|}
    \hline
    \textbf{} & \multicolumn{2}{c|}{\textbf{Quantum}} & \textbf{Classical} \\
    \hline
    $\{56o, 64e\}$ & Ion-trap & Neutral atoms & SHC \\
    \hline
    Runtime (days) & 1 & 0.73 & 7 \\
    \hline
    Physical qubits & $1.8$M & $758$ K & N/A \\
    \hline
    \end{tabular}
    \caption{The table shows the runtime estimates for GSEE with the studied quantum computations with ion-traps and neutral atoms versus the observed runtimes of a state-of-the-art classical method. The total physical qubit counts are reported.}
    \label{Table I} 
\end{table}

While the advances incorporated in our analysis are based in prior literature, it is important to stress here that the advances are combined in a highly non-trivial manner and their true impact can only be seen when rigorously combined. Importantly, the assessment of true quantum advantage comes from understanding full-stack performance of FTQC. Our work synthesizes these techniques into a cohesive, cross-layer framework that delivers almost 4 orders of magnitude reduction in runtime while keeping similar physical qubit counts. The integration itself provides insights on reducing the cost supported by extensive QRE analysis (see Sec.~\ref{subsec:detailed runtimes}). None of the innovations alone could bring down the runtime of quantum computation to just one day for $\{56o, 64e\}$ and just few days for the larger system sizes ($\{100o, 100e\}$ and $\{150o, 150e\}$), which is needed to claim the observed runtime quantum advantages (see Sec.~\ref{Sec:Classical_Benchmarking}).  All aspects of FTQC need to advance in parallel to have quantum computing realizing its potential and showcase a quantum advantage over classical for utility-scale applications with strong positive impact in our society. In that sense, our work puts forward evidence and concepts of accurate QRE in the service of realistically connecting quantum algorithms to positive real-world applications.

\subsection{Quantum Advantage } \label{Sec:Quantum_Advantage}

The quantum advantage the quantum computer is expected to have with our work is on accuracy of the energy estimation and runtime speed-up.

\subsubsection{Accuracy}~\label{Subsec:Accuracy}

As explained in Sec.~\ref{Sec:Impact}, classical methods fail to reliably converge within chemical accuracy of the energy estimation of the complex XVIII. In Table~\ref{Table XVIII classical}, we summarize all methods used in the literature (Density Function Theroy (DFT), Hartree-Fock (HF), Complete Active Space Self-Consistent Field (CASSCF), Density Matrix Renormalization Group (DMRG)) (see Ref.~\cite{complex_XVIII}) and in this work (SHCI) (see Sec.~\ref{Sec:Classical_Benchmarking} for more details). Importantly, all methods differ by more than 1 mHa, which is the target accuracy.

\begin{table}[h!]
    \centering
    \vspace{6pt}
    \begin{tabular}{|l|l|l|l|l|l|}
    \hline
    \textbf{Complex} & \textbf{DFT} & \textbf{HF}  & \textbf{CASSCF} & \textbf{DMRG}  & \textbf{SHCI (our work)} \\
    \hline
    XVIII & -3051.29506 & -7475.31480 & -7475.36738 & -7475.43923 & -7475.44040 \\
    \hline
    I & -2936.85035 & -7361.31568 & -7361.36040 & -7361.46138 & -7361.46465 \\
    \hline
    \end{tabular} 
    \caption{We report the electronic energies in Hartree for the complex I and complex XVIII $\{56o, 64e\}$ for the HF, CASSCF, DMRG methods as found in Ref.~\cite{complex_XVIII} and for the SHCI as calculated in this work.}
    \label{Table XVIII classical}
\end{table}

The classical methods in Table~\ref{Table XVIII classical} could be categorized according to how nearly exact methods they are. Given the CASSCF is only around 10 orbitals renders the energy estimation for the studied problem from Ref.~\cite{complex_XVIII} unreliable. Similarly, the the HF and DFT energies are unreliable, as in general they neglect electron correlations and thus, are not the optimal classical methods to use for the studied problem~\cite{complex_XVIII}. 

On the contrary, DMRG~\cite{chan2011density} and SHCI~\cite{Li2018} methods are both nearly exact methods and are comparable in performance. DMRG approximates the wavefunction as a matrix product state (MPS), while SHCI expands the wavefunction as a linear combination of determinants. If they both simultaneously converged to the same value with $<1$ mHa (or maybe even $<0.5$mHa), then we could be reasonably certain that we are predicting the true energy, since these two methods arrive at those results with very different approximations and computational strategies (see Sec.~\ref{Sec:Evidence_classical_benchmarking} for more details). Since they are not converged to this level (they differ by 1.2 mHa for complex XVIII $\{56o, 64e\}$), one could say that it is hard to know exactly what the error is -- SHCI could be converged and the DMRG results could be off, or the other way around. Usually, in classical quantum chemistry methods lower energy is assumed to be a better estimation, and therefore, it seems our calculation, with SHCI, could be considered more accurate.

\begin{figure}[h!]
    \centering
    \includegraphics[width=1.0\textwidth]{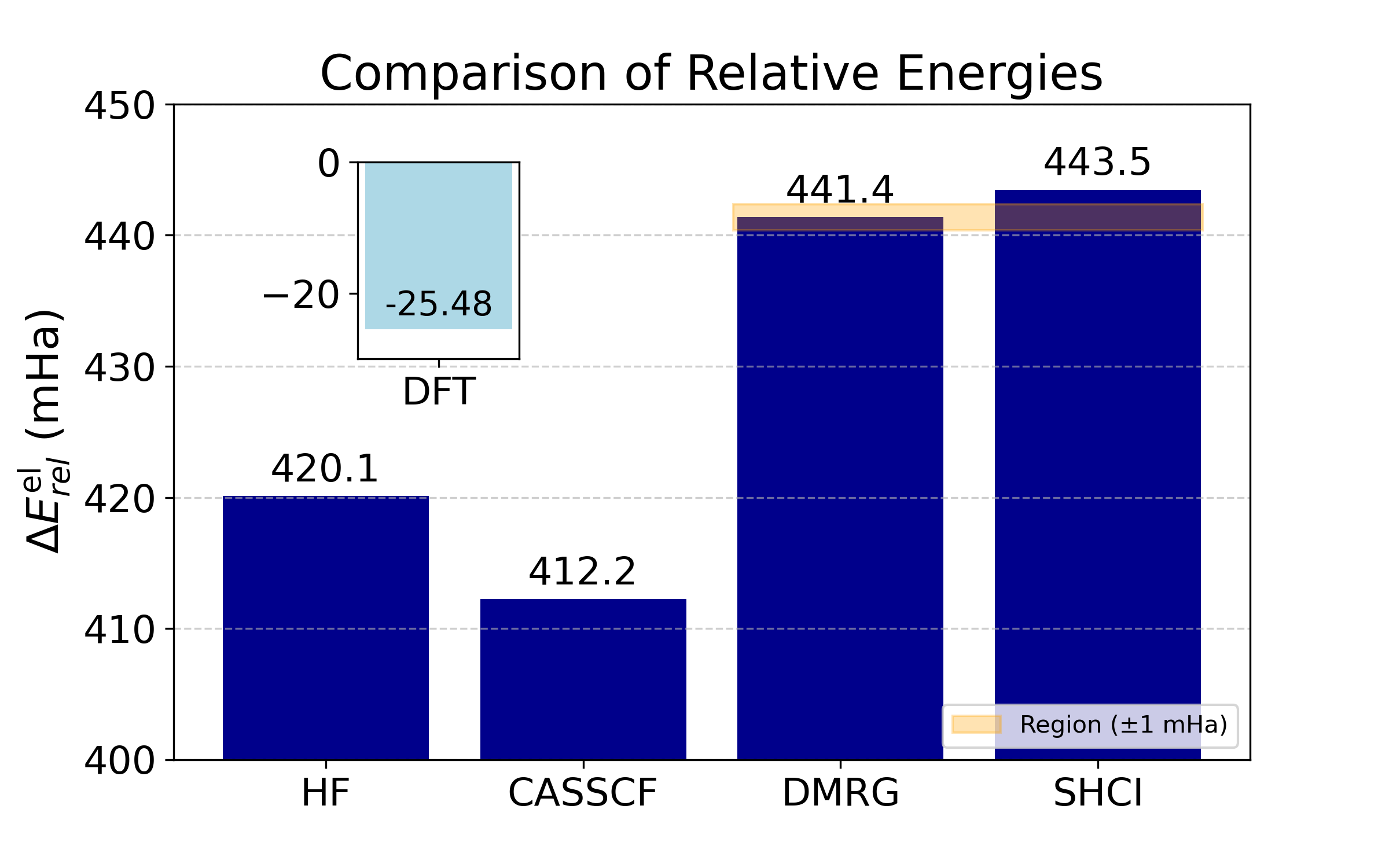}
    \caption{We plot the estimated reaction energy of Eq.~\eqref{Eq:DE} using the XVIII and I calculated classical energies from Table~\ref{Table XVIII classical}. For the rest of the systems involved in the reaction we used the energies reported in Table~\ref{Table Small molecules} (in Sec.~\ref{Sec:Evidence_classical_benchmarking}). For the explicit values of $\Delta E^{\mathrm{el}}_{\mathrm{XVIII}}
$ see Table~\ref{Table DE} in Sec.~\ref{Sec:Evidence_classical_benchmarking}. For better visibility, the DFT method is shown in the subplot in mHa. The region  $\pm$ 1mHa indicates that even the more closely aligned relative energies in absolute value still differ by more than the desired accuracy of 1mHa challenging the reliability of the methods. }
    \label{Fig.25-delta_E_XVIII.png}
\end{figure}

As explained in Sec.~\ref{Sec:Impact}, accurate energy estimations within chemical accuracy are necessary to better inform the chemical reactions in the catalytic cycles and the estimated reaction energy shown in Eq.~\eqref{Eq:DE}. In Fig.~\ref{Fig.25-delta_E_XVIII.png}, we show the calculated reaction energies of Eq.~\eqref{Eq:DE} for the methods discussed in Table~\ref{Table XVIII classical}, which include both estimations from the literature and also our own analysis. There we used the XVIII estimated classical energies from Table~\ref{Table XVIII classical} and the calculated energies of the rest of the systems (see Table~\ref{Table Small molecules} in Sec.~\ref{Sec:Evidence_classical_benchmarking} for a detailed discussion). The explicit values of $\Delta E^{\mathrm{el}}_{\mathrm{XVIII}}
$ can be found in Table~\ref{Table DE} in Sec.~\ref{Sec:Evidence_classical_benchmarking}. Interestingly,  DFT predicts the wrong sign, which means it is failing to even get the relative ordering correct. Moreover, DMRG and SHCI seem to have closely aligned relative energies still differ by more than $\pm$1 mHa. We also apply SHCI to increasing numbers of spin orbitals of the complex XVIII system (see Table~\ref{Table II} and Sec.~\ref{Sec:Classical_Benchmarking} for more details). 

The main analysis presented in this work focuses on achieving an energy error of $\epsilon = 1$ mHa—corresponding to better than chemical accuracy--which is the precision level to make meaningful predictions for the catalytic system under study. This accuracy threshold forms the core of our work and represents the critical regime in which quantum advantage becomes relevant for the problem at hand---especially given the uncertainty of classical computations in reliably verifying the energy errors of the methods used as already discussed. Contrary to classical methods, quantum algorithms can guarantee an analytical accuracy. In Fig.~\ref{Fig.18-accuracy_vs_orbitals} in App.~\ref{Sec:Evidence_quantum_advantage}, we explore the scaling of the quantum computational resources--both in terms of physical qubit count and runtime--as a function of increasingly tight energy error thresholds.

\subsubsection{Runtime}\label{subsec:runtime}

The runtime is evaluated using observed CPU-hours for classical computation (see Eq.~\eqref{CPU_equation} in Sec.~\ref{Sec:Classical_Benchmarking} and estimated QPU-hours for the quantum computation. The speed-up predicted in our work is due to cumulative innovations across all layers of FTQC--algorithmic, logical processor, quantum error correction, and physical hardware. While the use of a state-of-the-art algorithm contributes significantly, it is the full-stack co-design that enables us to reduce the rigorously estimated quantum runtime from 22 years~\cite{complex_XVIII} to just 1 day (see Table~\ref{Table I}). For comparison, the best classical method (SHCI) requires 7 days of total CPU-hours for the same problem size ($\{56o, 64e\}$) as discussed in Sec.~\ref{Sec:Classical_Benchmarking}.

\begin{table}[h!]
    \centering 
    \vspace{6pt}
    \begin{tabular}{|l|l|l|l|}
    \hline
    \textbf{} & \multicolumn{2}{c|}{\textbf{Quantum (estimated)}} & \textbf{Classical (observed)} \\
    \hline
     & \multicolumn{3}{c|}{\textbf{$\{56o, 64e\}$}}   \\
    \hline
    Runtime (days) & 1.0 & 0.73 & 7.0 \\
    \hline
    Physical qubits & $1.8$M & $758$ K & - \\
    \hline
     & \multicolumn{3}{c|}{\textbf{$\{100o, 100e\}$}}  \\
    \hline
    Runtime (days) & 4.8 & 1.7 & 27.8 \\
    \hline
    Physical qubits & $3.7$M & $1.96$ M & - \\
    \hline
     & \multicolumn{3}{c|}{\textbf{$\{150o, 150e\}$}}   \\
    \hline
    Runtime (days) & 8.7 & 3.1 & 294.4 \\
    \hline
    Physical qubits & $5.7$M & $3.1$ M & - \\
    \hline
    \end{tabular}
    \caption{The table shows the runtime estimates for GSEE with the studied quantum computations with ion-traps and neutral atoms versus the state-of-the-art classical SHCI method from our work for an increasing number of spin orbitals. The total physical qubit counts are also reported.}
    \label{Table II}
\end{table}

\begin{figure}[h!]
    \centering
    \includegraphics[width=0.8\textwidth]{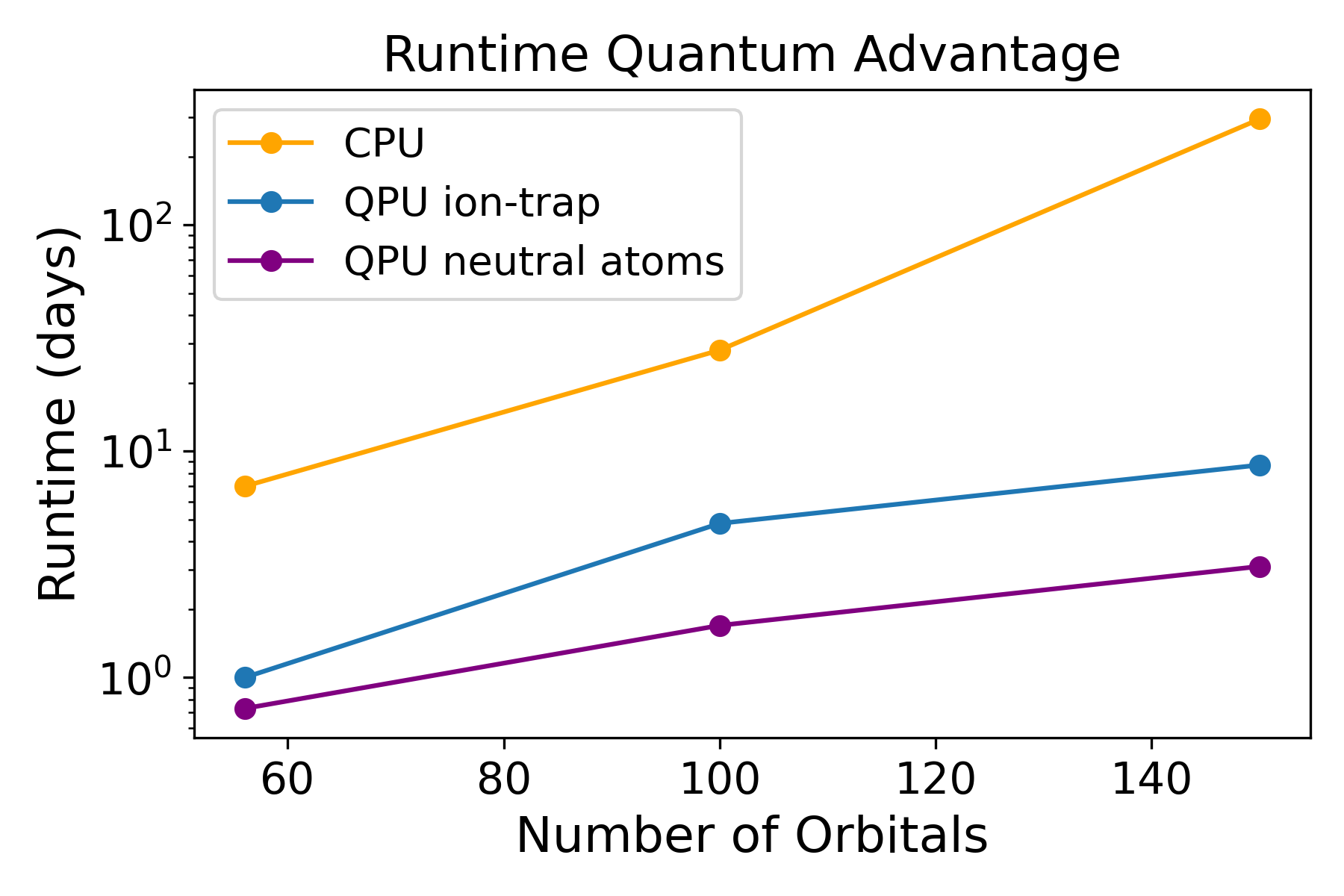}
    \caption{We plot the obtained CPU days of the SHCI method and QPU days for both the ion-trap and neutral atom quantum hardware modeling. This figure shows the observed runtime quantum advantage for the studied system sizes of the complex XVIII shown in Table~\ref{Table II}.  }
    \label{Fig.5-Runtime_vs_orbitals}
\end{figure}

Contrary to previous works -- both quantum and classical -- that restrict calculations to the $\{56o, 64e\}$ system, we extend the benchmarking to larger system sizes of $\{100o, 100e\}$ and \newline $\{150o, 150e\}$, using both quantum and classical methods (see Table~\ref{Table III}). This allows us to compare how runtime scales with the number of spin orbitals. In Table~\ref{Table III}, we report runtime estimates for two different quantum hardware modeling, i.e. ion-trap and neutral atoms. In Fig.~\ref{Fig.5-Runtime_vs_orbitals}, we plot the reported CPU and QPU runtimes to more clearly show the striking difference in the observed scaling with system size for classical versus the quantum method. Importantly, Fig.~\ref{Fig.5-Runtime_vs_orbitals} suggests that the asymptotic scaling of SHCI is exponential, while the asymptotic scaling of QPE is only polynomial given good initial states, which is the case for the studied systems (see App.~\ref{Sec:Evidence_quantum_advantage}).  

The underlying double-factorized tensor hyper-contraction with block-invariant symmetry shift and spectrum amplification (DFTHC+BLISS+SA) algorithm~\cite{DFTHC} used in this work has been benchmarked on molecular systems with up to $N = 150$ spin orbitals. The original study reports that DFTHC achieves chemically accurate representations with block-encoding Toffoli gate costs scaling as $\mathcal{O}(N^{0.96})$. Importantly, in our full-stack implementation—where this algorithm is compiled through all layers of fault-tolerant quantum computing (FTQC), including logical layout, error correction, and physical hardware modeling—we observe that the overall runtime scaling remains effectively linear with system size. This indicates that the overheads introduced by FTQC are well-managed and do not degrade the favorable asymptotic performance of the underlying algorithm. In App.~\ref{Sec:Evidence_quantum_advantage}, we elaborate in detail the innovations in the analysis of quantum algorithm performance that leads to the prediction of runtime quantum advantage.

\subsection{Classical Benchmarking } \label{Sec:Classical_Benchmarking}

\begin{figure}[h!]
    \centering
    \includegraphics[width=0.85\textwidth]{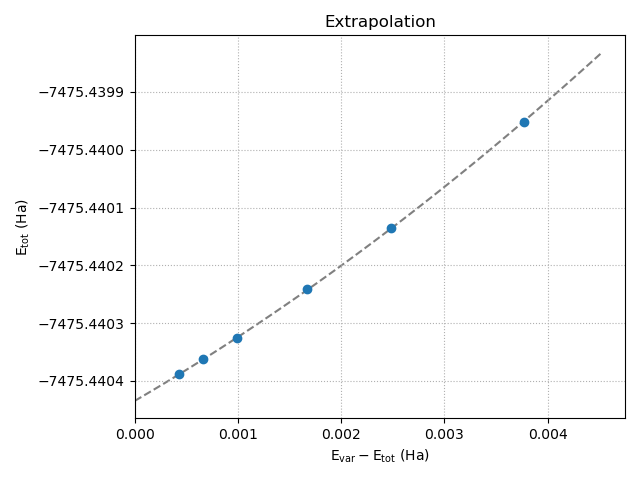}
    \caption{This plot shows the extrapolated energy to zero error of classical computation of the complex XVIII systems for $\{56o, 64e\}$ with the state-of-the-art method, SHCI. The heuristically observed uncertainty of the SHCI method for the $\{56o, 64e\}$ is $0.5$ mHa, but as explained in Sub-section~\ref{Subsec:Accuracy} the accuracy cannot reliably be validated. For $\{100o, 100e\}$ and $\{150o, 150e\}$, the relevant plots can be found in Sec.~\ref{Sec:Evidence_classical_benchmarking}.  }
    \label{Fig.6_Extrapolation}
\end{figure}

Classical computation methods of catalyst systems, like the complex XVIII, struggle to provide converged results on the energy estimation (see Sec.~\ref{Sec:Impact}). In this work, we utilize the Semistochastic Heat Bath Configuration Interaction (SHCI) method~\cite{Holmes2016, Li2018, Sharma2017}  a very fast selected configuration interaction plus perturbation theory method for obtaining almost exact Full Configuration Interaction (FCI) energies. From all the methods discussed in Sec.~\ref{Sec:Impact}, Selected configuration interaction has similar scaling to DMRG but potentially provides the best accuracy for the studied problem.
As shown in the Fig.~\ref{Fig.6_Extrapolation}, the SHCI results can be systematically improved by adding more configurations and extrapolating to zero error, yielding estimates of the total electronic energy $E_{tot}$ equal to $-7475.4404$ Ha .

Each point in Fig.~\ref{Fig.6_Extrapolation} has different constant factors of the wall-clock time of the classical solution, while the number of nodes and cores used remain the same, namely 8 nodes and 128 cores. As the x-axis value (i.e., the target energy error) decreases, the classical computational cost increases substantially. The leftmost point corresponds to the most demanding calculation, requiring 10 minutes on 8 nodes, each with 128 cores—amounting to 1024 cores in parallel and a total of 167 CPU-hours. In general, throughout this work, we report classical runtimes in CPU-hours, calculated as:

\begin{equation}\label{CPU_equation}
\text{CPU-hours} = \text{Wall-clock time (hours)} \times \text{Number of CPU cores}.
\end{equation}

This convention follows prior benchmarking methodology used to compare classical and quantum runtimes in resource estimation studies, including Ref.~\cite{goings2022reliably}, where wall-time multiplied by core count is used to quantify classical computational cost and enable a meaningful comparison to QPU time estimates under surface code compilation. Using the same core count with $\{56o, 64e\}$, we perform the classical calculations for the $\{100o, 100e\}$ and $\{150o, 150e\}$ systems and we find that they require approximately 668 and 7056 CPU-hours, respectively.

Moreover, as emphasized earlier, classical methods struggle to reliably achieve chemical accuracy at large system sizes as the one studied here for complex XVIII starting from 56 orbital and beyond. Classical methods in Ref.~\cite{complex_XVIII} notably omit uncertainty estimates for their reported energies.

In SHCI, the final energy is obtained by extrapolating results from multiple runs at different thresholds, using the perturbative correction \(\Delta E_{\text{pt}}\) as the extrapolation variable, where the total energy is given by \(E_v + \Delta E_{\text{pt}}\), with \(E_v\) representing the variational energy. The uncertainty in this extrapolation serves as an indicator of the final accuracy (refereed to as ``Case A''). A stronger measure of reliability is achieved when the perturbative correction is smaller than the threshold for chemical accuracy (refereed to as ``Case B''). An even more stringent and reliable criterion is when the perturbative correction is \emph{much smaller} than chemical accuracy, providing the highest level of confidence in the computed energy (refereed to as ``Case C'').

\begin{figure}[h!]
    \centering
    \includegraphics[width=0.8\textwidth]{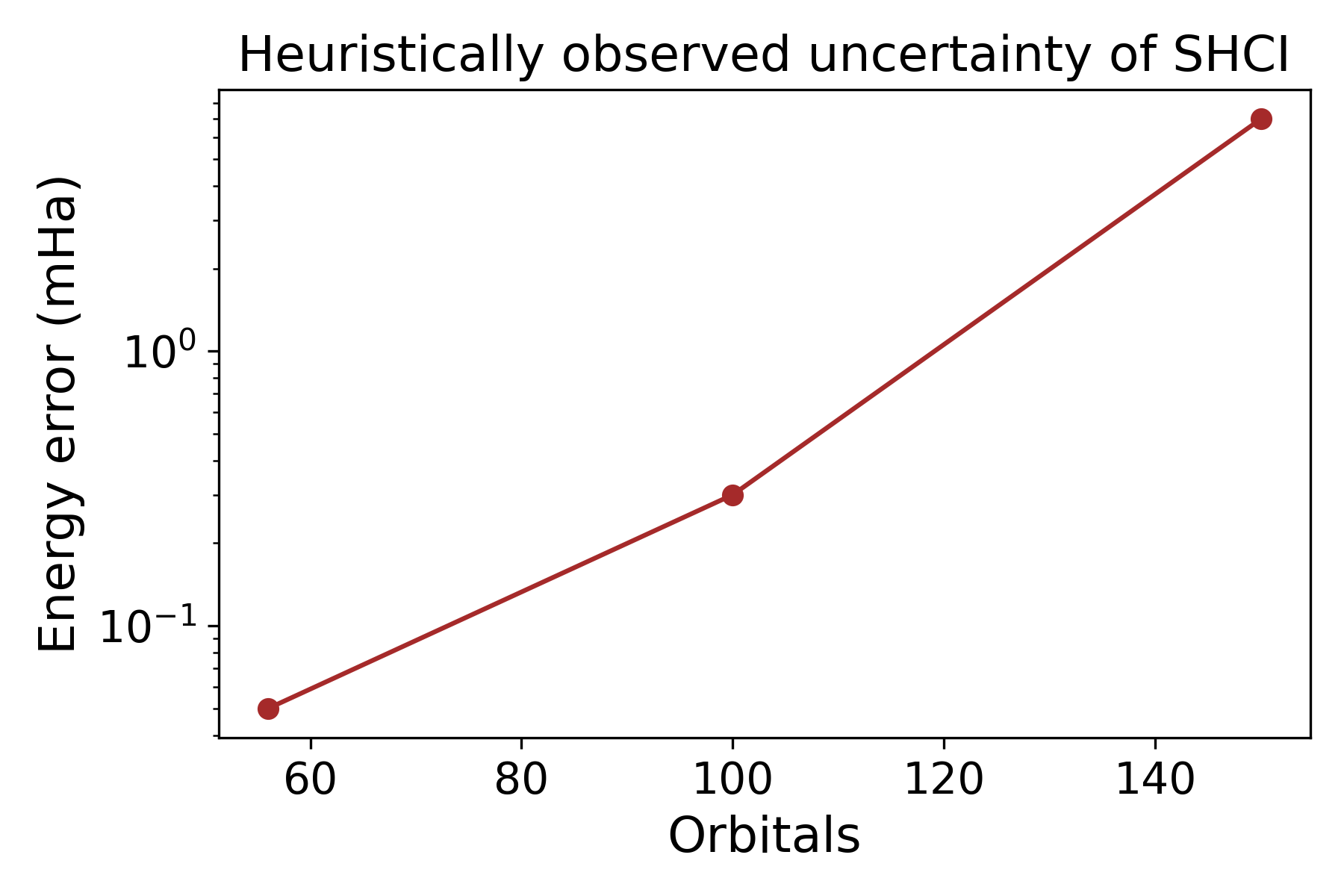}
    \caption{The heuristically observed uncertainty of SHCI with an increasing system size of the complex XVIII with increasing system size.}
    \label{Fig.20_classical_uncertainty}
\end{figure}

For the problem instance of $\{56o, 64e\}$, we use ``Case B'', while we use ``Case A'' for the larger two studies systems. These are earnest attempts on evaluating the energy errors associated with our classical solution on the estimated energy errors. We report that the  heuristically observed uncertainty for the $\{56o, 64e\}$, $\{100o, 100e\}$, and $\{150o, 150e\}$ systems are 0.05 mHa, 0.3 mHa, and 7 mHa, respectively. This suggests that the SHCI method (like any other classical method) exhibits growing uncertainty as system size increases (see Fig.~\ref{Fig.20_classical_uncertainty}).

We presented a classical benchmark using the SHCI method on the complex XVIII system. Among the classical approaches evaluated, SHCI stands out as the state-of-the-art in terms of energy estimations reported. Existing results from the literature serve as a reference point for classical intractability. Our study provides a strong foundation for estimating the boundary between classical feasibility and quantum advantage in catalytic systems.

\subsection{Viability}\label{Sec:Viability}

As already stressed in Sec.~\ref{Sec:Impact}, classical methods already fail to converge to a single-energy estimation within chemical accuracy for the $\{56o, 64e\}$ system. The SHCI method used here gives $-7475.4404$, while the DMRG gives $-7475.4392$ according to Ref.~\cite{complex_XVIII}. These two methods differ by 1.2 mHa, which is more than the desired accuracy of 1mHa. As explained in detail in previous sections, this uncertainty in energy estimations has significant implications for understanding both qualitatively and quantitatively the carbon transformation. Thus, already at problem sizes of around 50 orbitals, quantum computers could have a real-world positive impact. This is in accordance with the recent study on homogeneous catalysis systems~\cite{bellonzi2024}, where the authors suggest that at orbitals sizes of 50 or more, it will be feasible for fault-tolerant quantum computers to accelerate the discovery of homogeneous catalysts~\cite{bellonzi2024}.

\begin{figure}[h!]
    \centering
    \includegraphics[width=1.0\textwidth]{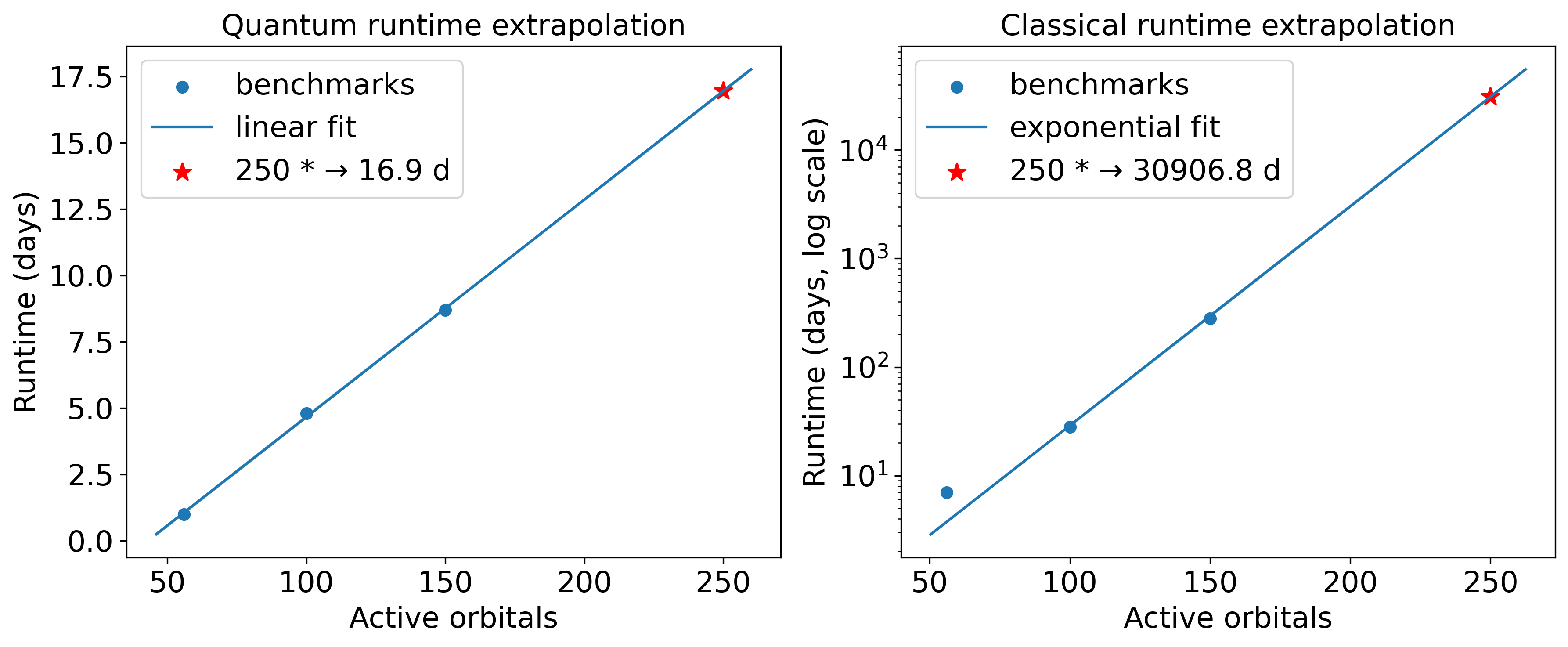}
    \caption{Linear regression of the runtime vs number of orbitals for the quantum methodology  (left) and the classical SHCI method (right) used in this work. We find a linear scaling for the quantum contrary to the exponential scaling for the classical method.}
    \label{Fig.7_Extrapolation_combined}
\end{figure}

Moreover, our analysis in Fig.~\ref{Fig.7_Extrapolation_combined} shows that the quantum computation scales linearly with system size (in terms of orbitals) contrary to classical that scales exponentially. The linear regression found is reported in Eq.~\eqref{Eq:quantum_regression}. This suggests that going to much larger systems sizes of 250 orbitals the runtimes are still very favorable for the quantum methodology discussed in this work with just around 17 days (see Fig.~\ref{Fig.7_Extrapolation_combined}). Even for 1000 orbitals, the extrapolated runtime is approximately 73 days. For such large system sizes, this is an acceptable value for runtime, especially given that such large number of orbitals can reliably provide a description of the true system,

\begin{equation}\label{Eq:quantum_regression}
\text{Runtime (days)} = 0.0818\,\text{orbitals} - 3.51.
\end{equation}

On the contrary, the classical methods (as already stressed earlier) scale exponentially with system size. In this work, we find that the extrapolated runtimes of the SHCI method for 250 and 1000 orbitals are $10^4$ and $10^{20}$ days, respectively (see Fig.~\ref{Fig.7_Extrapolation_combined}). These values are found by fitting the exponential model

\begin{equation}\label{Eq:classical_regression}
\text{Runtime (days)} = 0.02763
\exp\!\bigl(0.0465\,\text{orbitals}\bigr).
\end{equation}

Here we want to stress that we have compiled the quantum algorithm down to a detailed quantum hardware modeling of a distributed architecture for ion-trap systems by incorporating all significant advances on the logical processor and QEC layer of FTQC. Importantly, all these assumptions are based on realistic estimations of how quantum computation could be performed in the fault-tolerant era: 

\begin{itemize}
    \item \textbf{Quantum algorithm}: DFTHC+BLISS+DF~\cite{DFTHC} with proxy circuits from pyLIQTR~\cite{Obenland_pyLIQTR} that have the same number of T-gate counts as the original.
    \item \textbf{ZX calculus}: Compilation method from pyZX~\cite{kissinger2020Pyzx} to efficiently and reliably reduce the T-gate count, while implementing the same unitary with the original circuit.
    \item \textbf{Graph State Compilation}: Method implemented in Bench-Q~\cite{benchq} to efficiently lay out operations of a qasm circuit by mapping the circuits to graph states and leveraging the well studied field of graph theory.
    \item \textbf{Logical layout}: Effective all-to-all connectivity at the logical level that leverages the two-row logical layout of Elementary Logical Units (ELUs)~\cite{Simon} and the number of distinct two-qubit pairs of the proxy circuits.
    \item \textbf{QEC}: Magic State Cultivation (MSC)~\cite{MSC}to efficiently prepare good T-states with the minimal quantum resources needed.
    \item \textbf{Physical layout}: Detailed quantum hardware modeling of a distributed architecture~\cite{Simon} that takes into account the number of physical qubits that each ELU can realistically support and simulations on quantum hardware performance for the inter-ELU communications. 
\end{itemize}

This work builds on all the aforementioned realistic assumptions in all layers of the FTQC to tackle an application that poses significant challenges for classical computation. Our work provides strong evidence that all layers of computation are crucial for FTQC in the quest of quantum advantage. We presented a strong case that reduced the overall quantum computation runtime from 22 years to just 1 day, i.e. almost by 4 orders of magnitude. In the Appendix, (App.~\ref{Sec:Evidence_quantum_advantage}) we discuss in more detail all the aforementioned implementations and their viability. In addition, in Sec.~\ref{subsec:detailed runtimes} we present an in-depth QRE analysis that underscores the necessity of each innovation in enabling the reported performance improvements.

\begin{figure}[h!]
    \centering
    \includegraphics[width=1.0\textwidth]{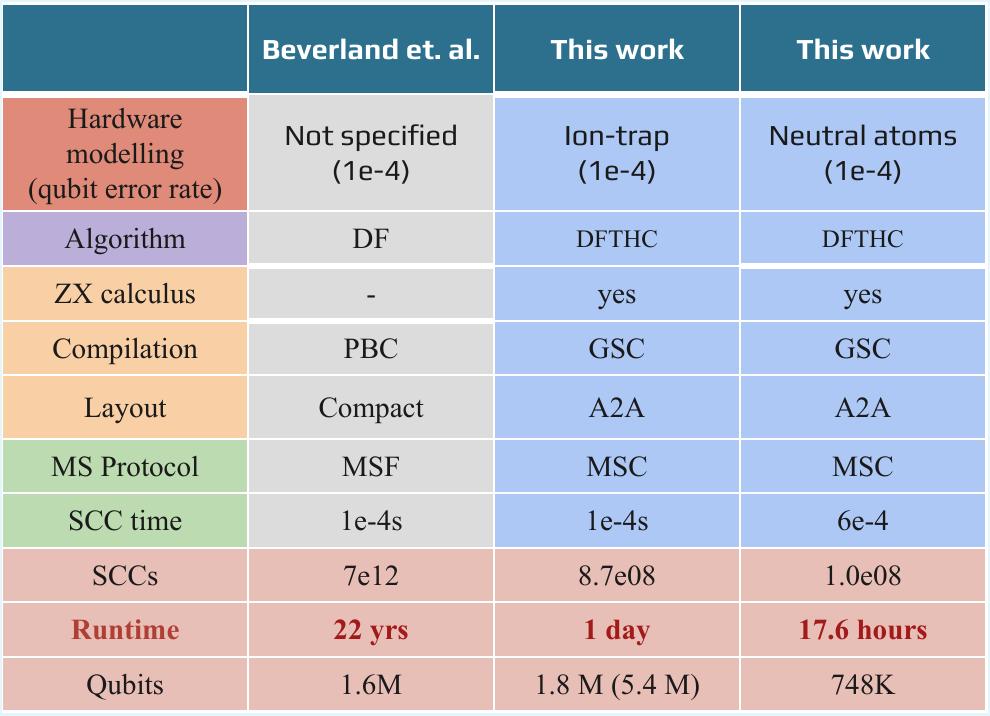}
    \caption{A short summary of the advances incorporated in this work compared to Beverland et. al.~\cite{complex_XVIII}. DF refers to double-factorized quantum algorithm discussed in~\cite{complex_XVIII}, while DFTHC refers to DFTHC+BLISS+SA quantum algorithm introduced in \cite{DFTHC}. PBC refers to Pauli-based measurement compilation scheme \cite{beverland2022}, while GSC is Graph State Compilation. The compact layout is introduced in \cite{GSC3}, while the A2A refers to an effective all-to-all based in \cite{Simon}. MSF refers to magic state factories discussed in \cite{Litinski2019MSF}, while MSC refers to magic state cultivation \cite{MSC}. For neutral atoms, the recent advances in erasure conversion~\cite{Wu2022ErasureConversion} were also included. SCC refers to surface code cycle time and SCCs to the total number of surface code cycles needed. A detailed discussion on all these advances and how we incorporated them can be found in App.~\ref{Sec:Evidence_quantum_advantage}. }
    \label{Fig.8-Detailed_results}
\end{figure}

Here we discuss how the advances incorporated in this work compare to Beverland et al.~\cite{complex_XVIII} (see Fig.~\ref{Fig.8-Detailed_results}). In our work we use proxy circuits contrary to Beverland et al. work~\cite{complex_XVIII} where only algorithmic logical counts were used. Importantly, even if we use explicit circuits of the double-factorized quantum algorithm~\cite{DF_pyLIQTR}, we still report a more than two orders of magnitude reduction in the runtime compared to what is reported in Beverland et al., namely going from 22 years down to around 58 days (see Sec.~\ref{subsec:detailed runtimes}). Here it is important to stress that we implemented a detailed quantum hardware modeling of a distributed architecture for ion-trap systems that can support SCC times of 1e-4 (see App.~\ref{Sec:Evidence_quantum_advantage} for details), leveraging the latest advances reported in the literature~\cite{Simon}. On the contrary, in Beverland et al.~\cite{complex_XVIII} the authors used SCC time of 6e-4 seconds and reported 130 years of runtime for trapped ion systems. Thus, a more strict comparison between the reported time of 130 years to 1 day results to more than four orders of magnitude speed-up.

Moreover, using QASM circuits in our work allow us to apply ZX calculus and fully leverage its capabilities. Our compilation scheme is also more detailed, providing explicit cycle estimates for graph state preparation, T measurements, and magic state distillation—in contrast to prior work, which focused solely on distillation. Additionally, we explore the impact of state-of-the-art quantum error correction (QEC) protocols and their effect on resource requirements. Finally, we incorporate a detailed quantum hardware model for ion-trap systems, taking advantage of an effective all-to-all connectivity at the logical level—a key strength of trapped-ion architectures. A detailed discussion on all these advances and how we incorporated them can be found in Sec.~\ref{subsec:detailed runtimes}.

Our work introduces concepts in the methodology of connecting  applications of positive real-world impact to quantum algorithms by incorporating a full-stack FTQC modeling. Our methodology goes beyond existing methods in the literature~\cite{beverland2022, bellonzi2024, scaling} and leads to compelling evidence of quantum advantage.

Although our implementation targets fault-tolerant quantum hardware, it is designed to be compatible with near-term FTQC hardware, making it feasible for the first generation of scalable quantum systems. For the representative case of $\{56o, 64e\}$, the required physical qubit counts are approximately 2 million for ion-trap architectures and 758 thousand for neutral atom platforms. These requirements fall within the publicly stated road-maps of leading quantum hardware companies, many of which project scaling to 2 million physical qubits by 2030~\cite{ionq2025roadmap}. Under the more detailed quantum hardware modeling for ion-trap architectures that explicitly incorporates inter-ELU communication costs, the physical qubit overhead rises to 5.4 million (see Sec.~\ref{subsec:detailed runtimes}). While this pushes implementation to a longer-term horizon, this estimate represents the first comprehensive qubit count that incorporates such realistic hardware constraints. We anticipate that future architectural innovations will significantly reduce this overhead.

Our resource estimates gives a strong predicted quantum advantage over both quantum and classical solutions previously demonstrated. Namely, for the $\{56o, 64e\}$, we achieve a $7.3 \times 10^3$ speed-up over previous quantum approaches, while keeping the physical qubit counts approximately the same. Over classical methods, we report a 6.25× speed-up, while also offering more reliable energy estimations within the desired accuracy. Moreover, our work offers a linear scaling with system size contrary to the exponential scaling of classical methods. The implications of the studied problem to carbon absorption and green energy set our approach as a technical innovation, but also a solution with significant real-world impact. Importantly, our methodology could be implemented to other problems formulated for FTQC~\cite{alexeev2025perspectivequantumcomputingapplications, otten2024quantumresourcesrequiredbinding, nguyen2025quantumcomputingcorrosionresistantmaterials, nguyen2025quantumcomputingcorrosionresistantmaterials} and could also become compatible with other quantum hardware platforms.

\subsection{Detailed breakdown of Quantum Resource Estimates}\label{subsec:detailed runtimes}

Here we discuss in more detail how each innovation contributes to the runtime and physical qubit count estimation. It is hard to completely isolate the contribution of each innovation to the physical estimates (runtime and qubits), but we explore the effect of each layer of FTQC to the extent possible. Importantly, this analysis further stress the importance of our contribution of carefully incorporating realistic advancements of all aspects of FTQC and leveraging the maturity of the field to show how to target utility-scale applications on future quantum hardware. 

\begin{figure}[h!]
    \centering
    \includegraphics[width=1.0\textwidth]{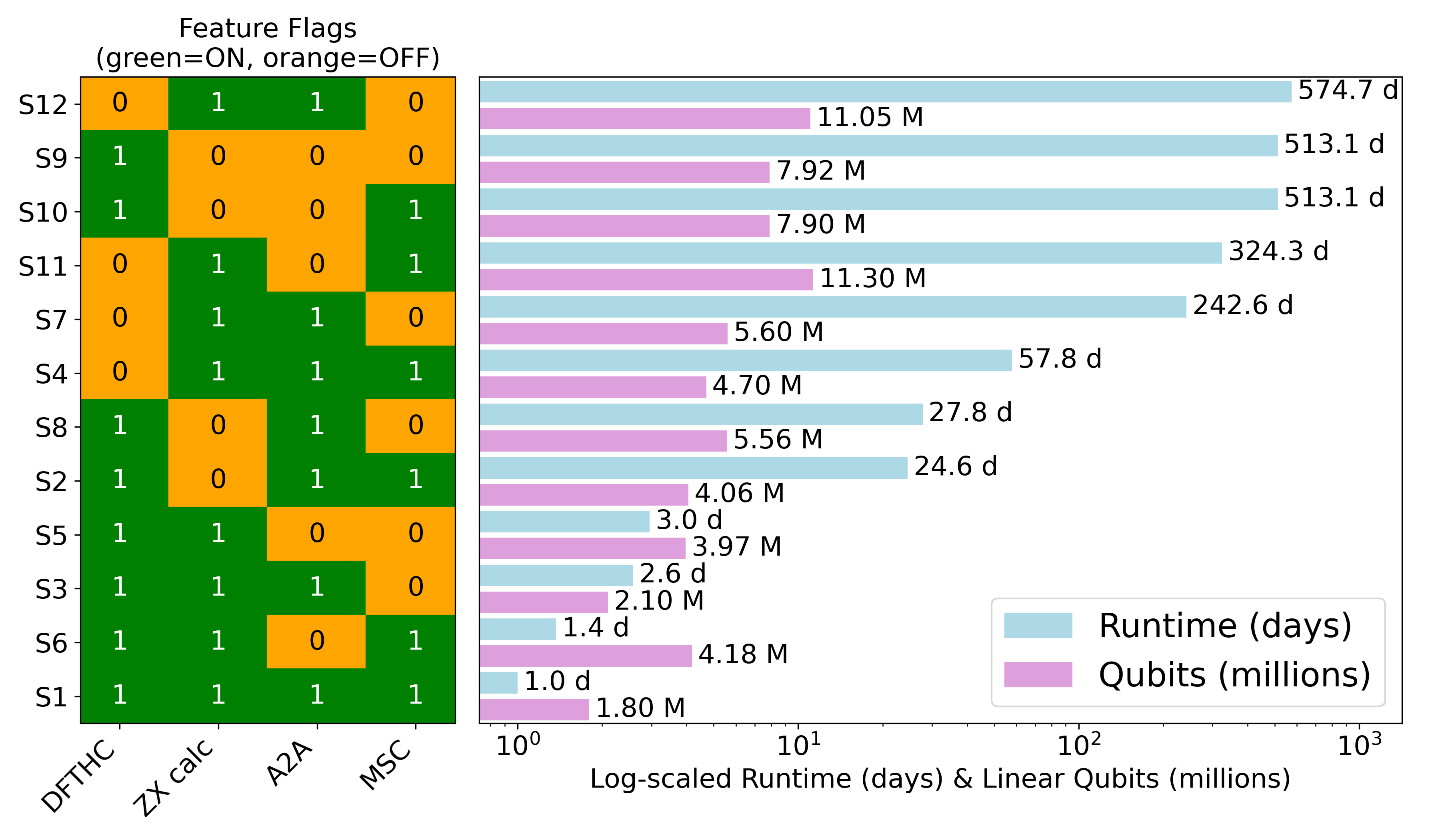}
    \caption{This figure show how the incorporation of different features affect the overall runtime of the quantum computation. On the left, we have a binary feature map where 1 and 0 (green and orange colors, respectively) indicate the innovation is included or not included. On the right, we have the corresponding runtime qubit estimates of the given simulation. We numerate the simulations from $S1$ to $S12$ and order them from largest to lowest runtime value. The physical qubits correspond to the basic hardware modeling.}
    \label{Fig.28-Flags}
\end{figure}

We numerate the simulations from $S1$ to $S12$ where we have a different combination of the innovations already discussed. In Fig.~\ref{Fig.28-Flags}, we have a binary feature map where 1 and 0 (green and orange colors, respectively) indicate the innovation is included or not included. Here we discuss in more detail what this entails:

\begin{itemize}
    \item \textbf{DFTHC}: This refers to the quantum algorithmic level:
    \begin{itemize}
        \item on (green) means that the proxy QASM circuits from pyLIQTR~\cite{Obenland_pyLIQTR} for the \newline DFTHC+BLISS+SA~\cite{DFTHC} are used
        \item off (orange) means that the explicit QASM circuits from  pyLIQTR~\cite{Obenland_pyLIQTR} on the double-factorized quantum algorithm~\cite{DF_pyLIQTR} are used.
    \end{itemize} 
    \item \textbf{ZX calculus}: ZX calculus can efficiently and reliably reduce the T-gate count, while implementing the same unitary with the original circuit.
    \begin{itemize}
        \item on (green) means that the ZX calculus~\cite{kissinger2020Pyzx, Duncan2019ZX} is applied to the QASM circuits
        \item off (orange) means that the ZX calculus is not applied to the QASM circuits
    \end{itemize}
    \item \textbf{A2A}: A2A refers to the effective all-to-all connectivity at the logical level that leverages the two-row logical layout of ELUs~\cite{Simon} and the number of distinct two-qubit pairs of the proxy circuits.
    \begin{itemize}
        \item on (green) means that the effective all-to-all is used
        \item off (orange) means that a simple two-row bus architecture is used instead.
    \end{itemize}
    \item \textbf{MSC}: MSC refers to the QEC methodology used. 
        \begin{itemize}
        \item on (green) means that the magic state cultivation~\cite{MSC} was used
        \item off (orange) means that magic state distillation~\cite{Litinski2019MSF} was used instead.
    \end{itemize} 
\end{itemize}

We would like to stress here that optimization of the hyper-parameters of the graph state preparation is also an important task. This can result to efficiently laying out operations of a circuit by mapping circuits to graph states and leveraging the well studied field of graph theory where different optimizations could be implemented. In this work, we performed an extensive optimization on the hyper-parameters and identified the following: teleportation threshold = 4, min neighbor degree = 4, max num neighbors to search = 1e6, use fully optimized dag = True, teleportation distance = 2.

\begin{figure}[h!]
    \centering
    \includegraphics[width=1.05\textwidth]{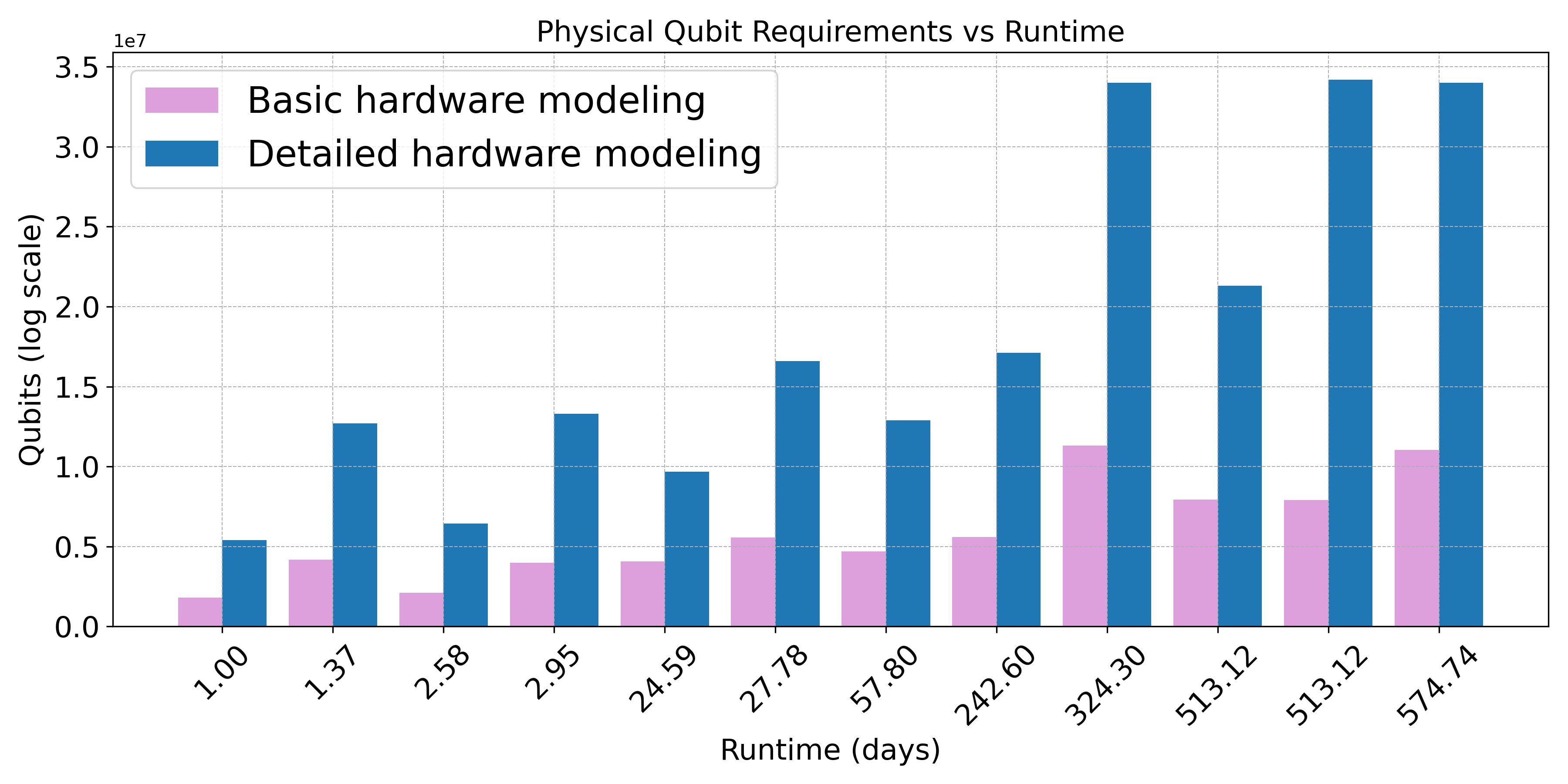}
    \caption{We compare the number of physical qubits needed for the basic vs the detailed hardware modeling for the simulations reported in Fig.~\ref{Fig.28-Flags}.}
    \label{Fig.Detail_vs_basic}
\end{figure}

In Fig.~\ref{Fig.Detail_vs_basic}, we compare the number of physical qubits needed for the basic versus the detailed hardware modeling for the simulations reported in Fig.~\ref{Fig.28-Flags}. The basic quantum hardware modeling is closer to what presented in the work of Ref.~\cite{beverland2022} and does not incorporate the inter-ELU communication. On the contrary, the detailed quantum hardware modeling incorporates a feasible inter-ELU modular architecture as extensively discussed in App.~\ref{subsec:Hardware}. The inflation we observe in the number of physical qubits with the detailed quantum hardware modeling is expected and are in accordance with the literature. For example, the recent work~\cite{scaling} that discuss a detailed superconducting quantum modeling result to higher physical qubit counts from what was reported for relevant problems in Beverland et al.~\cite{beverland2022}. 

The incorporation of detailed quantum hardware modeling for the distributed architecture further underscores the significance of our work, as it demonstrates a feasible path toward implementing the application on a quantum computer in terms of both runtime and physical qubit requirements. Hardware companies road-maps show a milestone of 2M physical qubits by 2030~\cite{ionq2025roadmap}. This suggests that the detailed quantum hardware modeling will already be feasible in the upcoming years. Importantly, in future work we plan to explore how to reduce the number of physical qubits needed for this specific application. Moreover, our analysis could be extended to other applications that could potentially be mapped to less than 2M physical qubits and we plan to further explore this.

\begin{table}[h!]
\centering
\begin{tabular}{|c|c|c|c|c|c|c|}
\hline
\textbf{Simulation} & \textbf{Runtime (days)} & \textbf{Qubits (M)} & \textbf{DFTHC} & \textbf{ZX calculus} & \textbf{A2A} & \textbf{MSC} \\
\hline
S1 & 1.0   & $1.8$   & 1 & 1 & 1 & 1 \\
\hline
S9 & 513.1 & $7.92$ & 1 & 0 & 0 & 0 \\
\hline
S4 & 57.8  & $4.7$  & 0 & 1 & 1 & 1 \\

\hline
\end{tabular}
\caption{Comparison of Simulations S1, S4, and S9 in terms of runtime, qubit count, and enabled features as shown in Fig.~\ref{Fig.28-Flags}. Importantly, the features ZX calculus, A2A and MSC combined have a larger impact in both the runtime and number of qubits.}
\label{tab:sim_comparison}
\end{table}

Next, we want to highlight the importance of combining all features to reduce the runtime to just one day. As shown in Table~\ref{tab:sim_comparison}, applying only the algorithmic improvements results in a runtime of approximately 1.4 years. In contrast, enabling all other features—All2All connectivity, ZX calculus, and MSC—while keeping DFTHC disabled, reduces the runtime more substantially to 57.8 days. Although each of these innovations contributes meaningfully on its own, analyzing them in isolation does not provide a robust case for quantum advantage. 

Moreover, the impact of different combination of these innovations can be highly unintuitive in general. For example, going from 82 years down to 8 months by only enabling ZX calculus (S13 to S7 in Table~\ref{tab:reductions}), suggests a 123x runtime reduction. While only enabling the DFTHC algorithm instead (S13 to S8 in Table~\ref{tab:reductions}), we have a 1054x runtime reduction. Thus, the projected runtime reduction by combining both (enabling DFTHC and ZX caclulus) would be 1.3e05, calculated as the product of the independent speed-ups (namely, 123x and 1054x from ZX and DFTHC, respectively). Instead, when ZX and DFTHC are enabled we have a reduction of 1.1e04, which is less than the product of the independent reductions. It is interesting to note though, that the runtime reduction of ZX calculus and DFTHC is comparable once the two-row logical architecture is used (see Table~\ref{tab:reductions-two-row}) with ZX calculus resulting to approximately 1.6x speed-up compared to DFTHC. This is another clear indication of the complexity of the runtime impact of different innovations and the importance of our work. 

\begin{table}[h!]
\centering
\begin{tabular}{|c|c|c|c|c|c|c|}
\hline
\textbf{Simulation} & \textbf{Runtime} & \textbf{Qubits (M)} & \textbf{DFTHC} & \textbf{ZX calculus} & \textbf{A2A} & \textbf{MSC} \\
\hline
S13 & 82 years  & $8.9$   & 0 & 0 & 1 & 0 \\
\hline
S7 & 8 months & $5.6$ & 0 & 1 & 1 & 0 \\
\hline
S8 & 27.8 days & $5.56$ & 1 & 0 & 1 & 0 \\
\hline
S3 & 2.6 days  &  $2.1$  & 1 & 1 & 1 & 0 \\
\hline
\end{tabular}
\caption{Comparison of different simulations in terms of runtime, qubit count, and enabled features to understand the runtime reductions from enabling DFTHC and ZX calculus while the all-to-all logical layout is used.}
\label{tab:reductions}
\end{table}

\begin{table}[h!]
\centering
\begin{tabular}{|c|c|c|c|c|c|c|}
\hline
\textbf{Simulation} & \textbf{Runtime} & \textbf{Qubits (M)} & \textbf{DFTHC} & \textbf{ZX calculus} & \textbf{A2A} & \textbf{MSC} \\
\hline
S10 & 513.1 & $7.9$ & 1 & 0 & 0 & 1 \\
\hline
S11 & 324.3 & $11.3$ & 0 & 1 & 0 & 1 \\
\hline
S6 & 1.4  &  $4.18$  & 1 & 1 & 0 & 1 \\
\hline
\end{tabular}
\caption{Comparison of different simulations in terms of runtime, qubit count, and enabled features to understand the runtime reductions from enabling DFTHC and ZX calculus while the two-row logical layout is used instead of the all-to-all.}
\label{tab:reductions-two-row}
\end{table}

Our simulations demonstrate that it is the careful combination of these techniques that enables a compelling and defensible argument for scalable quantum advantage. Fig.~\ref{Fig.28-Flags} and Fig.~\ref{Fig.Detail_vs_basic} strongly support our claims that no single innovation is sufficient to reduce the runtime to just 1 day while maintaining a feasible number of physical qubits. Similarly, Table~\ref{tab:sim_comparison}, \ref{tab:reductions} and \ref{tab:reductions-two-row} highlight the importance of carefully evaluating the collective impact of different advances on FTQC stack. The true effect of any single technique can often be counterintuitive—and, in some cases, even misleading—if evaluated in isolation. 

\begin{figure}[h!]
    \centering    \includegraphics[width=1.05\textwidth]{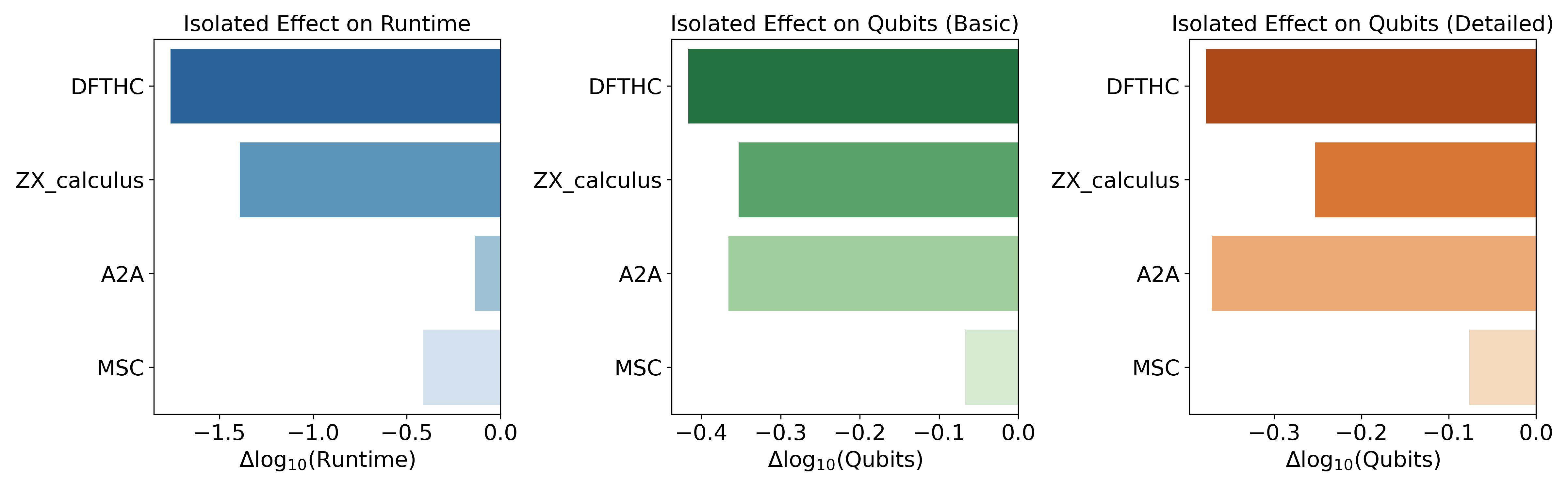}
    \caption{We study the isolated effect of each key feature on the runtime and number of physical qubit counts for the basic and detailed quantum hardware modeling.}
    \label{Fig.effects}
\end{figure}

In Fig.~\ref{Fig.effects} we present an analysis of the effect of the each key feature enabled in the FTQC stack to better understand its impact on the runtime and number of physical qubits. The methodology used was to compare the runtime and physical qubit counts to the baseline case S1, where all features were enabled. Then, only one feature is turned on at a time to explore the impact in the key metrics studied here.

The analysis presented here further underscores the critical role of our methodology, which disentangles complex inter-dependencies and enables informed co-design of FTQC across the stack. Rather than relying on vague or complexity-theoretic arguments to justify quantum advantage, we rigorously explore the interplay of innovations by fully compiling a realistic application instance to a fault-tolerant quantum architecture. Identifying and coherently integrating these advances is a non-trivial undertaking that demands a deep, cross-disciplinary understanding of the FTQC stack. Most importantly, our work provides strong, evidence-based support for the feasibility of the observed quantum advantage under plausible assumptions on every aspect of FTQC stack.

\section{Conclusions}\label{Sec:Conclusions}

We demonstrated that utility-scale applications could be achieved on FTQC with greatly reduced resources via full-stack co-design of FTQC, realized via detailed and accurate Quantum Resource Estimations analysis. The focus of our work was on ion-trap systems that have orders of magnitude slower gate times compared to other platforms. Our work highlights how crucial all layers of quantum computation are in the quest of quantum advantage. By incorporating key innovations in all layers of computation (algorithmic, logical, quantum error correction and physical), we reduced the predicted runtime for the calculation of ground state energies for large catalyst systems to just one day compared to years reported at the previous state-of-the-art work~\cite{beverland2022}. 

Our work suggests that quantum computers could realistically tackle several such systems per year and therefore help elucidate the mechanisms around carbon transformation to methanol. Importantly, our methodology could be applied to other applications (beyond ground state energy estimation) and even to other hardware systems (as we show with neutral atom hardware). Moreover, we carefully assess the classical bottlenecks for the studied problem and provide classical estimations with the SHCI method to help define the boundary of quantum advantage up to 150 orbitals. Importantly, our work uses proxy circuits for large system sizes and go beyond the common practice of relying solely on logical counts (qubits and T-gates) to get estimations on the runtime and physical qubit counts for the quantum computation. We reduce the runtime estimation almost by 4 orders of magnitude compared to previous work while also incorporating a detailed distributed quantum hardware modeling through photonic interconnects.

Our work utilizes three different open-source software tools, namely pyLIQTR, Bench-Q and pyZX. The integration of the tools is challenging and further work would be beneficial in that direction. Moreover, further development of certain features in each tool could be addressed in future work. For example, a detailed distributed architecture modeling for other quantum hardware systems could be incorporated. The replacement of the DFTHC+BLISS+SA proxy circuits with explicit circuits in pyLIQTR. We believe the final estimations on runtime and physical qubits will be similar, but it will be beneficial for increasing the accuracy of the methodology. Additionally, the explicit circuits could further help with the co-design of FTQC full-stack development. 

Our work shows how utility-scale applications of ground state energy estimations of large catalyst systems that face significant bottlenecks for classical methods could realistically and practically be tackled on a FTQC. Performing accurate ground state energy estimations of large catalyst systems is the key to understanding the mechanisms that govern the efficient transformation of carbon dioxide to methanol through Ru-based catalyst intermediate states. Understanding this mechanism would represent a major leap forward in a carbon-neutral chemical economy given that green hydrogen is used. This is because methanol is considered a renewable fuel that could directly be used in combustion engines or power generation.

Our work focus on the methodology of connecting quantum algorithms to applications of positive impact in our society and provides a systematic analysis of quantum algorithm performance that leads to compelling evidence of achievable quantum advantage.

\section{Acknowledgments}
We would like to thank Peter D. Johnson for scientific discussions at an earlier stage of this work.

\appendix
\section{Appendix} \label{Sec:Evidence_quantum_advantage}

Here we discuss in more detail the implementations (see Fig.~\ref{Fig.8-Detailed_results}) across the FTQC stack of our work that gives the observed quantum advantage claimed in Sec.~\ref{Sec:Impact}. First, we focus on ion-trap systems, which constitute the main subject of our analysis. We then comment on neutral-atom systems as a complementary study, illustrating that the methodology developed here could potentially be refined for other hardware platforms. Similarly, our analysis could be extended to a broader set of applications.

\subsection{Ion-traps}

\subsubsection{Algorithmic layer} As already mentioned earlier, we use the DFTHC quantum algorithm recently introduced in \cite{DFTHC}. Specifically, we use pyLIQTR to generate proxy circuits for the complex XVIII with different system sizes that have approximately the same number of logical qubits and T-gate counts compared to the resource analysis given in \cite{DFTHC}. In more detail, we have 994 logical qubit and 8.2e09 T-gate count compared to 924  logical qubit and 8.2e08 T-gate count given in \cite{DFTHC}. In future work, we plan to have explicit circuits from the DFTHC quantum algorithm~\cite{DFTHC}, but given that the proxy circuits are based on double-factorized block encodings we believe that the current approach is a good proxy of the original circuits. This is because the main overhead from the computation comes from the T-gate count, which is exactly the same in our implementation as in the paper. Importantly, we have also included the number of block-encoding circuit repetitions needed $ \frac{\pi \lambda}{2\sigma_{\mathrm{PEA}}}$~\cite{DFTHC}, where $\sigma_{\mathrm{PEA}}$ is the target accuracy of 1mHa. For more details, see Table~\ref{Table III}.

\begin{table}[h!]
    \centering
    \vspace{6pt}
    \begin{tabular}{|l|l|l|l|l|}
    \hline
    \textbf{} & \textbf{proxy circuits} & \textbf{original circuits} \\
    \hline
    56o - logical qubits & 994 & 924  \\
    \hline
    56o - T gates & 8.2e08 & 8.2e08 \\
    \hline
    100o - logical qubits & 1872 & 1960  \\
    \hline
    100o - T gates & 4.24e09 & 4.24e09 \\
    \hline
    150o - logical qubits & 2954 & 2870  \\
    \hline
    150o - T gates & 1.12e10 & 1.12e10 \\
    \hline
    \end{tabular}
    \caption{Comparison of the logical qubits and the T-counts between the proxy circuits generated in pyLIQTR with the estimated counts in \cite{DFTHC}, referred to as original circuits. }
    \label{Table III} 
\end{table}

Here we would like to stress that even if we focus our analysis to the double-factorized quantum algorithm, i.e. the explicit QASM quantum circuits from pyLIQTR~\cite{DF_pyLIQTR} instead of the proxy DFTHC circuits, then the overall runtime of the computation is around $58$ days. Even in this case, our methodology results in more than two orders of magnitude runtime speed-up, namely $1.4e02$, compared to the previous work~\cite{complex_XVIII} of 22 years. If we compare with the reported 130 years (Ref.~\cite{complex_XVIII}), then we have almost three orders of magnitude runtime speed-up, namely $8e02$.

\subsubsection{ZX calculus}

We applied ZX calculus on the proxy circuits to reduce the number of T-gate counts used by using the open-source library PyZX~\cite{kissinger2020Pyzx}. The quantum circuit is represented as a ZX-diagram and a phase teleportation simplification strategy can be applied to reduce the T-count~\cite{kissinger2019reducing, Duncan2019ZX}. This technique allows non-Clifford phases to combine and cancel by propagating non-locally through the quantum circuit while keeping the optimized and original circuits are equal~\cite{kissinger2019reducing}. In our circuits, we observed a 8.6x reduction in T-gate counts, specifically from 9e08 down to 1.6e08. This reduction is in accordance with what is expected for such large circuits~\cite{kissinger2019reducing, kissinger2020Pyzx}. It is important to note here though that certain operations (such as reset and phase gates) are skipped and therefore, future work could implement this and further study the effect of ZX calculus to the full circuits.  In subsection ~\ref{Subsec:resource_reductions}, we discuss in more detail the effect of ZX calculus in combination with the compilation scheme.

\subsubsection{Compilation}

\begin{figure}[h!]
    \centering
    \includegraphics[width=0.6\textwidth]{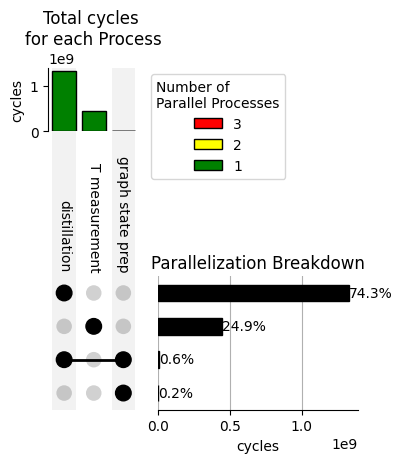}
    \caption{Resource breakdown for the QEC-layer surface code cycles for the effective ``all-to-all'' connectivity at the logical level. This figure shows the number of surface code cycles used by each of the three primary logical architecture operations (entanglement, distillation, and T-state to T-gate).}
    \label{Fig.9-QEC_cycle-allocation}
\end{figure}

For the logical quantum processor layer of computation, we have used and optimized the Graph State Compilation (GSC) scheme~\cite{GSC3} in Bench-Q~\cite{benchq}. GSC methodology is based on compiling the circuits to graph states and exploiting the many optimization schemes available~\cite{GSC3, GSC2, GSP1}. In the Bench-Q~\cite{benchq} implementation of GSC, the total cycles needed are equal to the cycles for T-measurement, distillation, and graph state preparation. In Fig.~\ref{Fig.9-QEC_cycle-allocation}, we plot the parallelization breakdown of the three different cycles involved.

Moreover, GSC implementation in Bench-Q~\cite{benchq} has a number of hyper-parameters related to the graph state preparation~\cite{GSC3, GSC2, GSP1} and we have optimized the hyper-parameters to allow for an optimal set. For a naive implementation of the GSC, the total runtime of the quantum computation can significantly increase, i.e. by 25 times and the number of physical qubit count of the basic quantum hardware modeling can almost double. 

As stressed throughout, it is the combination of distinct advances that enables the significant overall runtime reduction. For example, applying ZX calculus simplification strategy on the logical circuits has significant implications for the compilation scheme with a 8.6 reduction of the T-measurement cycles and a factor of 2 reduction in the total cycles needed for distillation. Thus, we see a significant reduction of 1.3e03 in the total cycles needed for graph state compilation. This is in accordance with what is expected since the total preparation tocks increase at least quadratically with the interaction-graph density~\cite{GSP1, GSC2, GSC3}.

Moreover, it is important to note here that the graph state preparation cycles needed before and after the implementation of pyZX changes significantly. As mentioned earlier, the studied circuits have certain simplifications in their gate operations, and it seems that the effect of pyZX is due not only to the ZX calculus but also to the significant cost reduction of the "graph state preparation" part in the total cycles allocation because of the much simplified entangled states that need to be prepared. After pyZX, the total cycles are governed by "distillation" and "T measurement" which is the main overhead of other compilation schemes, such as Pauli-based compilation.  This might suggest that such compilation schemes might be more resource efficient than graph state compilation. Future work is needed to further assess and disentangle the cause of the runtime savings that is observed here to ZX calculus and potentially to other compilation schemes.

\subsubsection{Logical Layout}

\begin{figure}[h!]
    \centering
    \includegraphics[width=\textwidth]{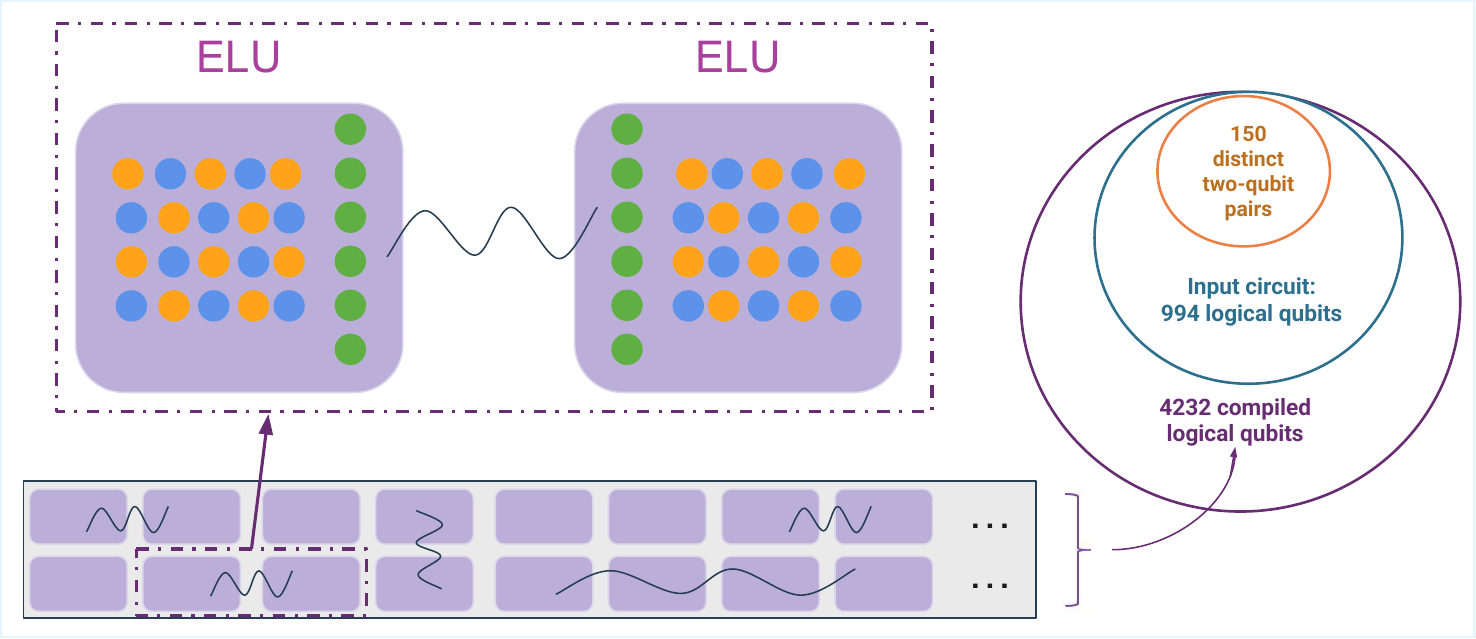}
    \caption{Here we present the detailed quantum hardware modeling with effective all-to-all connectivity that supports the inter-ELU communication at the logical level. Two separate ELUs built from a collection of trapped ions, where most of of the ions in each trap are used to encode a logical surface code qubit (depicted with blue and orange colors). The rest (depicted with green) are used to collect and refine entanglement to be used for a two-qubit lattice surgery operation that enables communication between two distinct ELUs. For the problem instance of $\{56o, 64e\}$, 4232 total ELUs are the logical qubit overhead at the compiled layout (depicted in the outer purple circle). The effective all-to-all connectivity is feasible given that approximately just 150 distinct two-qubit interaction pairs (orange) exist at the algorithmic circuit (blue) with 994 logical qubits. }
    \label{Fig.10-inter-ELU}
\end{figure}

In the analysis that we have performed, we have incorporated a detailed ion-trap quantum hardware modeling of a modular architecture based on the work of Devitt et al. ~\cite{Simon}. As discussed there, it is natural for ion-trap systems to form a single crystal which could be an Elementary Logical Unit (ELU) and host 30 logical qubits with all-to-all connectivity. But, this is not enough for encoding utility-scale problems of thousands of logical qubits. Devitt et al.~\cite{Simon} proposed a realistic model of multiple crystals chained together to form a segmented linear trap of ELUs (see Fig.~\ref{Fig.10-inter-ELU} for an illustration). 

Each ELU consists of memory, computational and communication ions and supports one logical qubit and two-qubit gate operations between distinct ELUs (see Fig.~\ref{Fig.10-inter-ELU}) which the latter are performed through a purification protocol introduced in Devitt et al.~\cite{Simon}. This model uses realistic hardware assumptions on ion-ion entanglement probability of 2.18e-4 and required targeted fidelities for operation at 0.99 for the performance of the operations in two distinct ELUs~\cite{Simon}. According to Monroe et al. each ELU could realistically hold around 1000 ions~\cite{Monroe_1K, ref22-Simon, ref23-Simon}. 

Effective all-to-all connectivity at the logical level is a key assumption in our analysis for minimizing the physical qubit overhead. While IonQ’s latest roadmap~\cite{ionq2025roadmap} suggests that true all-to-all connectivity may be achievable for systems with up to 2 million physical qubits and thousands of logical qubits, the practical realization of such connectivity remains uncertain. In this section, we explore how a logical layout could be implemented in practice and present evidence that full all-to-all connectivity may not be strictly necessary. Instead, we argue that even an effective all-to-all scheme based on modular architectures with few sparse long range connections supported by swap operations~\cite{Simon} or by a photonic router outside of the ELUs~\cite{Brown2016npjqi16034}. Therefore, our analysis covers both the optimistic projections in which true all-to-all connectivity is achievable, and the more cautious view that such connectivity may be difficult to implement at scale.

For the problem instance of $\{56o, 64e\}$ we start with around 1000 logical qubits at the algorithmic level which are then mapped to 4232 compiled logical qubits. Each compiled logical qubit is encoded in one ELU. This is in accordance with how the graph state compilation scheme performs~\cite{GSC3, GSC2, GSP1}. Importantly, the distinct number of logical pairs needed from our proxy circuit at the algorithmic level is just 150 (see illustration in Fig.~\ref{Fig.10-inter-ELU}). This suggests that at the logical level, we do not need to have true all-to-all connectivity, but instead an effective all-to-all (see Fig.~\ref{Fig.10-inter-ELU}) that allows the fraction of the necessary compiled logical qubits to communicate with photonic links. The ELUs are laid out in a two-row design (see Fig.~\ref{Fig.10-inter-ELU})~\cite{GSC3} and even with optimal compilation few sparse long-range interactions could remain. However, we anticipate that the number of such interactions will be minimal, and thus their overall impact on runtime could potentially be negligible. Alternatively, this challenge could be mitigated by assuming the availability of sufficient remote entangled Bell pairs, for example, distributed via a photonic router external to the ELUs~\cite{Simon, Brown2016npjqi16034}. Importantly, even if we drop the all-to-all advancements incorporated, the runtime is only slightly increased to 1.3 days, while the number of physical qubits needed is doubled (see Sec.~\ref{subsec:detailed runtimes}).

\subsubsection{QEC layer}

A key assumption in Devitt et al.~\cite{Simon} is that the surface code cycle time determines the overall runtime of the computation at the inter-ELU level. For this to be possible, a certain number of communication ions is used to keep the runtime of inter-ELU operation the same as an inner-ELU operation~\cite{Simon}. Therefore, a trade-off between the number of communication ions and the speed of SCC time exists. For the slower SCC runtimes for ions (which is well within what is possible) we have SCC time equal to 3e-3 second and the number of communication ions per ELU in the order of hundreds. For faster SCC times of 1e-4 s (which is still possible with today’s technology~\cite{Simon}) we have an increased number of ions per ELU in the order of thousands. Later in this section, we discuss in more detail the inter-ELU entanglement in ion-trap hardware (see App.~\ref{subsec:entanglement}).

Regarding QEC strategies, we explore both the effect of magic state distillation~\cite{Litinski2019MSF}, but also the more recent work on magic state cultivation~\cite{MSC}. Magic state cultivation is possible in the case of complex XVIII with a logical volume at 7.6e11 if we take into account that the end-to-end distilled magic state error rates from cultivation is at 4e-12 or smaller, which comes from an extrapolation of the performance of cultivation protocol at noise strength of 1e-4~\cite{MSC}. In Ref.~\cite{MSC}, they perform simulation for the end-to-end error rates at 4e-11 given 5e-4 noise strength. This suggests that 4e-12 end-to-end distilled magic state error rates should be possible for 1e-4 noise strength. Moreover, the MSC protocol requires approximately 460 physical qubits per factory and its most resource-intensive stage—the escape phase—necessitates 20 parallel attempts~\cite{MSC}. To remain consistent with the time-optimal scenario, we assume all 20 attempts run in parallel. As a result, the total physical qubit requirement per MSC factory is 9200 (i.e., $20$ attempts $ \times 460$ qubits).

Importantly, even if we instead use magic state distillation~\cite{Litinski2019MSF}, the total runtime of the quantum computation increase to just 2 days and the number of physical qubits is slightly increased for both the basic and detailed hardware modeling (see subsection~\ref{subsec:detailed runtimes}). Here we would also like to stress that for the larger system sizes (namely, $\{100o, 100e\}$ and $\{150o, 150e\}$) only magic state distillation was used.

\subsubsection{Hardware modeling}\label{subsec:Hardware}

We use in detail the quantum hardware modeling introduced by Devitt et al.~\cite{Simon} of the modular architecture at the inter-ELU level which considers the entanglement generation between spatially separated ELUs. As a representative example of the detailed modular architecture modeling incorporated, we use the inter-ELU entanglement model~\cite{Simon} (see App.~\ref{subsec:entanglement} for details) and update the relevant physical parameters based on the most recent experimental data reported by Monroe et al.(see App.~\ref{subsec:entanglement}). This allows us to more accurately assess entanglement success probabilities under realistic conditions. Furthermore, we explicitly analyze the trade-off between the surface code cycle (SCC) runtime and the number of communication ions required per ELU to meet the entanglement rate target~\cite{Simon}.

\begin{figure}[h!]
    \centering
    \includegraphics[width=0.8\textwidth]{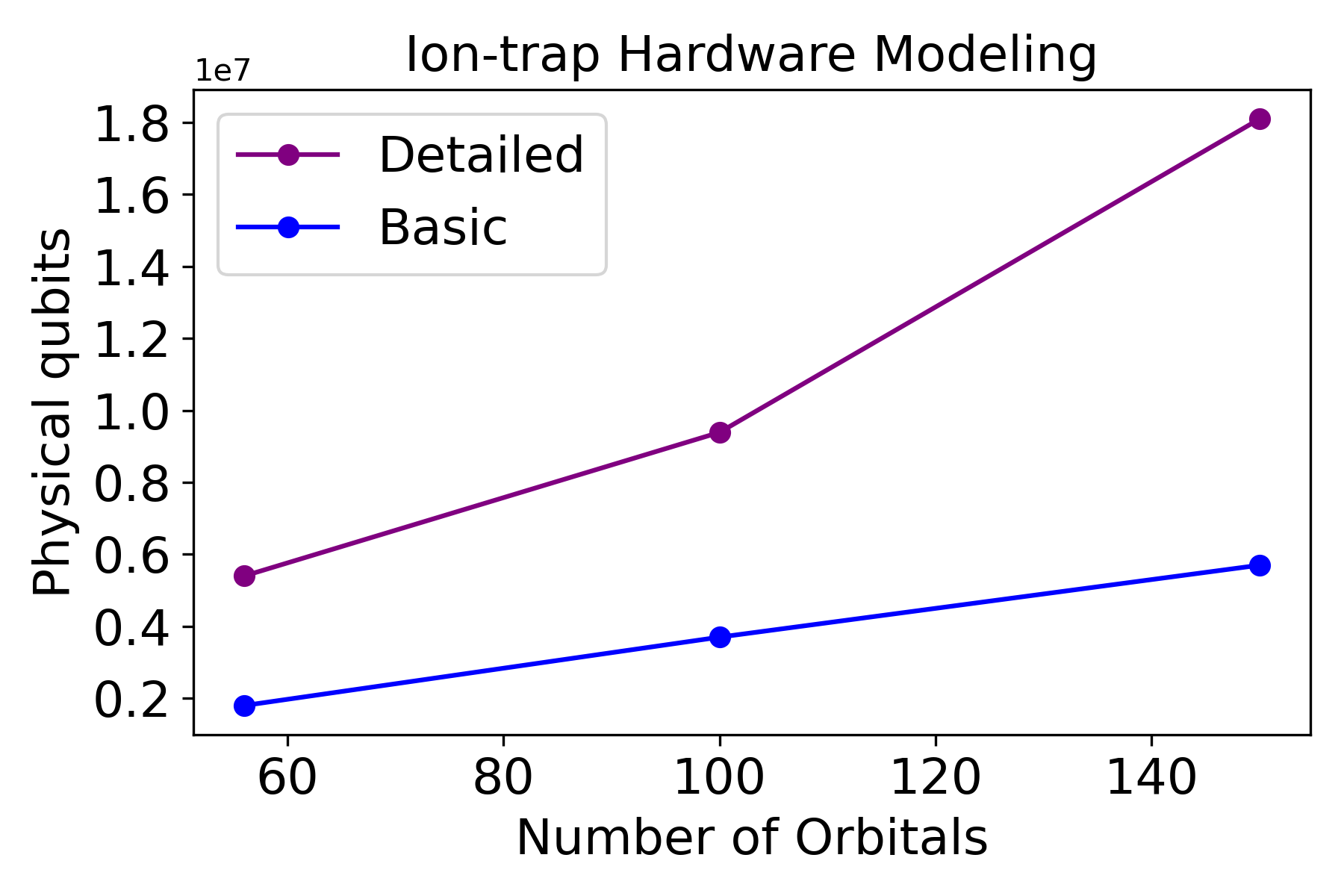}
    \caption{Number of physical qubits needed based on the basic ion-trap architecture modeling and detailed ion-trap modular architecture modeling~\cite{Simon}. The overhead of the physical qubits in the detailed hardware modeling is coming from the number of ions needed per ELU (see Fig.~\ref{Fig.13-ELU}) to support the SCC runtime of 1e-4 seconds at the inter-ELU level~\cite{Simon}. }
    \label{Fig.17-Qubits_vs_orbitals}
\end{figure}

\begin{figure}[h!]
    \centering
    \includegraphics[width=1.0\textwidth]{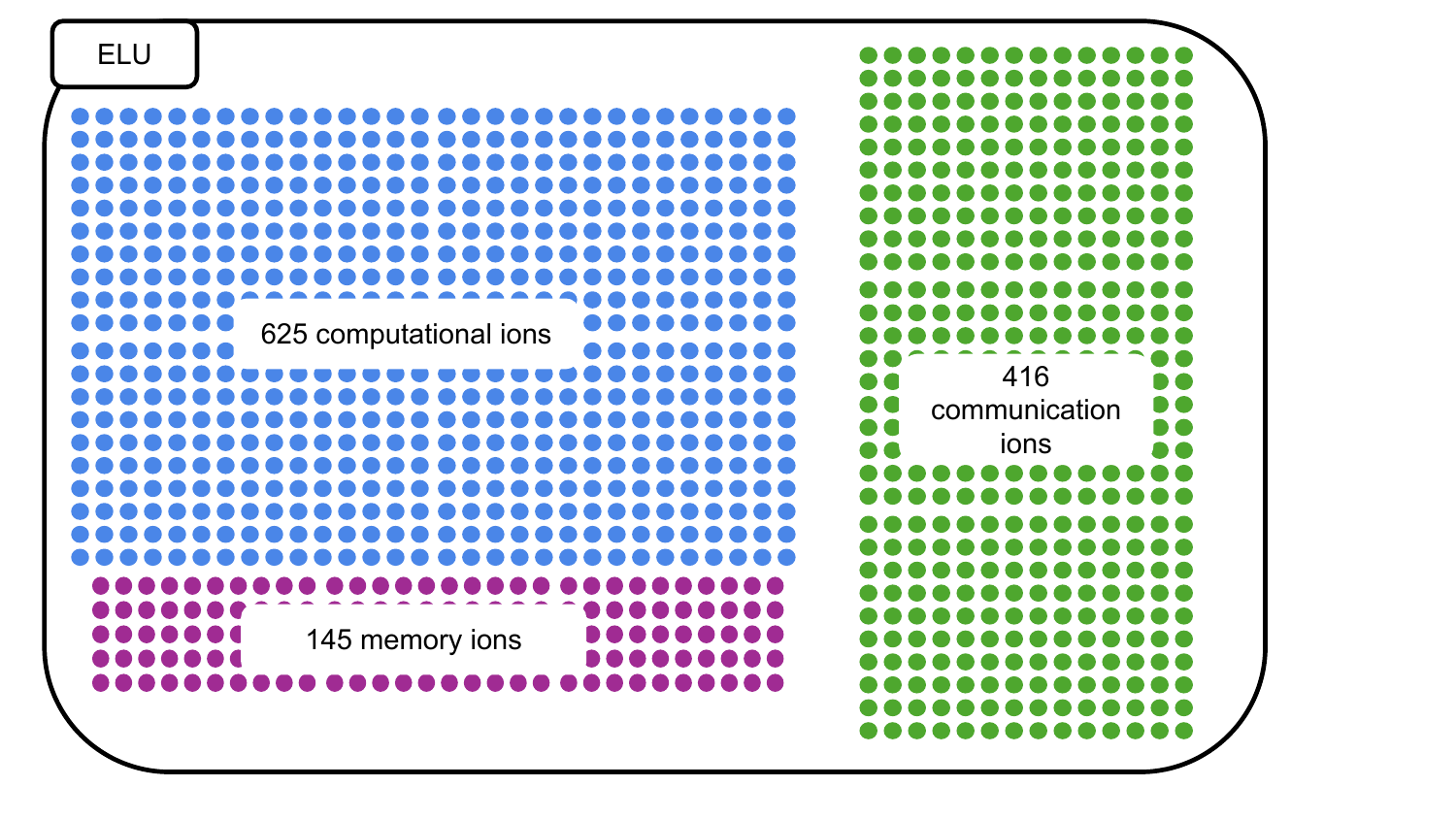}
    \caption{In each ELU for the problem instance of $\{56o, 64e\}$, we have 1186 ions split into 416 communications, 625 computational and 145 memory ions.}
    \label{Fig.13-ELU}
\end{figure}

\begin{table}[h!]
    \centering 
    \vspace{6pt}
    \begin{tabular}{|l|l|l|l|l|l|}
    \hline
    \textbf{Complex XVIII} & \multicolumn{3}{c|}{\textbf{Number of ions per ELU}} &  \multicolumn{2}{c|}{\textbf{Number of ELUs}} \\
    \hline
    \textbf{System size} & {\textbf{Communication}} & \textbf{Computation} & \textbf{Memory} & \textbf{Data} & \textbf{Factory} \\
    \hline
    $\{56o, 64e\}$ & 416 & 625 & 145 & 4232 & 38 \\
    \hline
    $\{100o, 100e\}$  & 473 & 841 & 167 & 7817 & 12  \\
    \hline
    $\{150o, 150e\}$ & 473 & 841 & 167 & 12215 & 14 \\
    \hline
    \end{tabular}
    \caption{Detailed breakdown of the number of communication, computation and memory ions per ELU for an increasing system size. The number of date and factory ELUs are also reported. For the 56 orbitals, the factories correspond to magic state cultivation, while for the other systems to distillation.}
    \label{Table V}
\end{table}

We find that 1.8M physical qubits are needed in a basic quantum hardware modeling that does not take into account the number of ions that each ELU contains to support the 1e-4s SCC time at the inter-ELU level~\cite{Simon}. Once this is accounted for in the detailed quantum hardware modeling of a modular architecture, the quantum computation requires 5.4M physical qubits (see Fig.~\ref{Fig.17-Qubits_vs_orbitals}). The overhead is coming from the number of ions needed per ELU to support the SCC runtime of 1e-4 and a total runtime of the quantum computation at 1 day. In each ELU, we have 1186 ions split into 416 communications, 625 computational and 145 memory ions (see Fig.~\ref{Fig.13-ELU}) and with a total number of 4350 data ELUs and 54 cultivation ELUs for the case of $\{56o, 64e\}$. In Table~\ref{Table V}, we present the breakdown of ions per ELU of the complex XVIII for $\{100o, 100e\}$ and $\{150o, 150e\}$ system sizes.

\subsection{Neutral atoms}

Next, we discuss the complementary study using neutral-atoms quantum hardware modeling. The algorithmic, ZX calculus and GSC methodology is the same as for ion-traps. The two-row logical layout is still a good proxy for neutral atoms as recently discussed in Ref.~\cite{tworow_NA}. In future work, we plan to extend the hardware modeling of neutral atoms to a more detailed version of the inner- and inter- ELU operations. Regarding QEC, we implemented and used a scheme to support transversal surface code for neutral atoms following the work of Ref.~\cite{sunami2025transversal}. Importantly, erasure conversion enabled by high-fidelity atom loss detection in neutral atom arrays significantly raises the effective error threshold to 4.15e-2~\cite{Wu2022ErasureConversion} from 1.3e-2 threshold used in conventional surface code approximations. These implementations result in 17.6 hours of quantum computation runtime with 748K physical qubits (see Fig.~\ref{Fig.8-Detailed_results}).

Our work showcases how accurate and realistic QRE analysis could play a meaningful and impactful role in the co-design of quantum hardware modeling. The methodology builds on a modular QRE stack that allows to efficiently access the impact and consistency of different innovations in all layers of computation.

\subsection{Quantum computation runtime with explicit overlap values}

\begin{figure}[h!]
    \centering
    \includegraphics[width=0.8\textwidth]{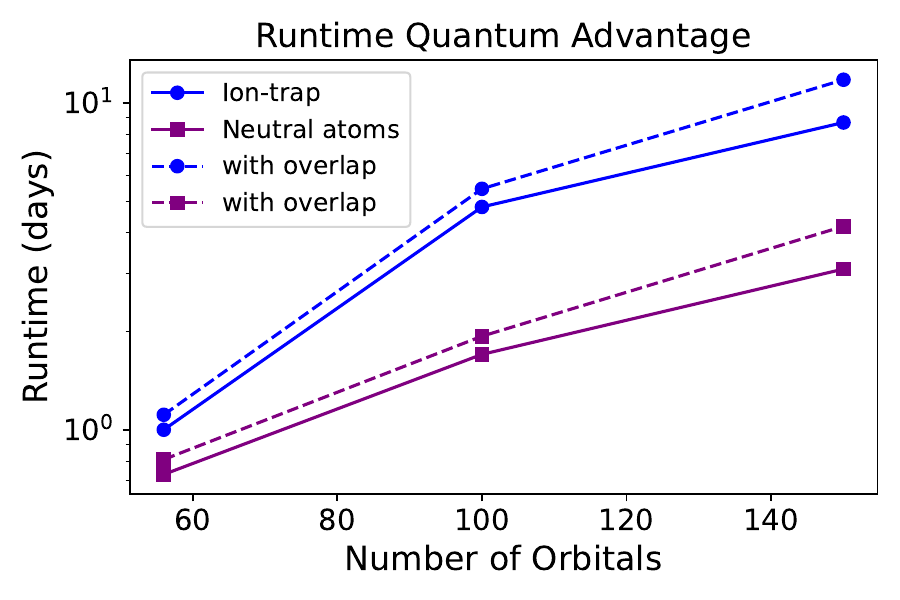}
    \caption{We compare the QPU runtimes for ion-trap and neutral atom systems with (dotted line) and without (continuous line) the overlap values. This plot suggests that the observed runtime quantum advantage holds for the studied system sizes of the complex XVIII even when explicit overlap values are incorporated in the runtime calculation.}
    \label{Fig.16-Runtime_vs_orbitals_with_overlap}
\end{figure}

The quantum computation speed-ups over classical still remains approximately the same even if we include the overhead of the ground state preparation (see Fig.~\ref{Fig.16-Runtime_vs_orbitals_with_overlap} and Table~\ref{Table:overlaps}). In that case, the runtime is increased by a factor $1/\gamma$, where $\gamma$ is the overlap. The SHCI method in our work results to relatively high overlap values for all system sizes studied (see Table~\ref{Table:overlaps}), and therefore, we  only observe a slight increase in the runtimes (see Fig.~\ref{Fig.16-Runtime_vs_orbitals_with_overlap}). Moreover, we still observe a linear scaling of the quantum computation runtime (see Fig.~\ref{Fig.16-Runtime_vs_orbitals_with_overlap}) in accordance with what is expected for acceptable values of the overlap~\cite{postoponing}. There are plenty other methods to use that could potentially increase the overlap value if needed, and while the reported values seem to be sufficient~\cite{Gratsea2024, gratsea2022rejectGSP}, future work could further explore how the overlap values could to be increased.

\begin{table}[h!]
    \centering
    \vspace{6pt}
    \begin{tabular}{|l|l|l|l|l|}
    \hline
    \textbf{} & \multicolumn{2}{c|}{\textbf{Quantum Runtime} (days)} & \textbf{Classical Runtime} (days) & \textbf{Overlap} \\
    \hline
    System size & Ion-trap & Neutral atoms & SHCI (this work) & SHCI (this work) \\
    \hline
    $\{56o, 64e\}$ & 1.1 & 0.8 & 7.0 & 0.91 \\
    \hline
    $\{100o, 100e\}$ & 6.4 & 2.3 & 27.8 & 0.88\\
    \hline
    $\{150o, 150e\}$ & 11.8 & 4.2 & 294.4 & 0.74 \\
    \hline
    \end{tabular}
    \caption{The table shows the runtime estimates for the studied systems for both quantum and the classical state-of-the-art method. For the quantum runtimes, the overhead from the overlap values is included.}
    \label{Table:overlaps}
\end{table}

\subsection{Scaling of the quantum computational resources vs accuracy}\label{Subsec:Scaling_vs_accuracy}

\begin{figure}[h!]
    \centering
    \includegraphics[width=1.0\textwidth]{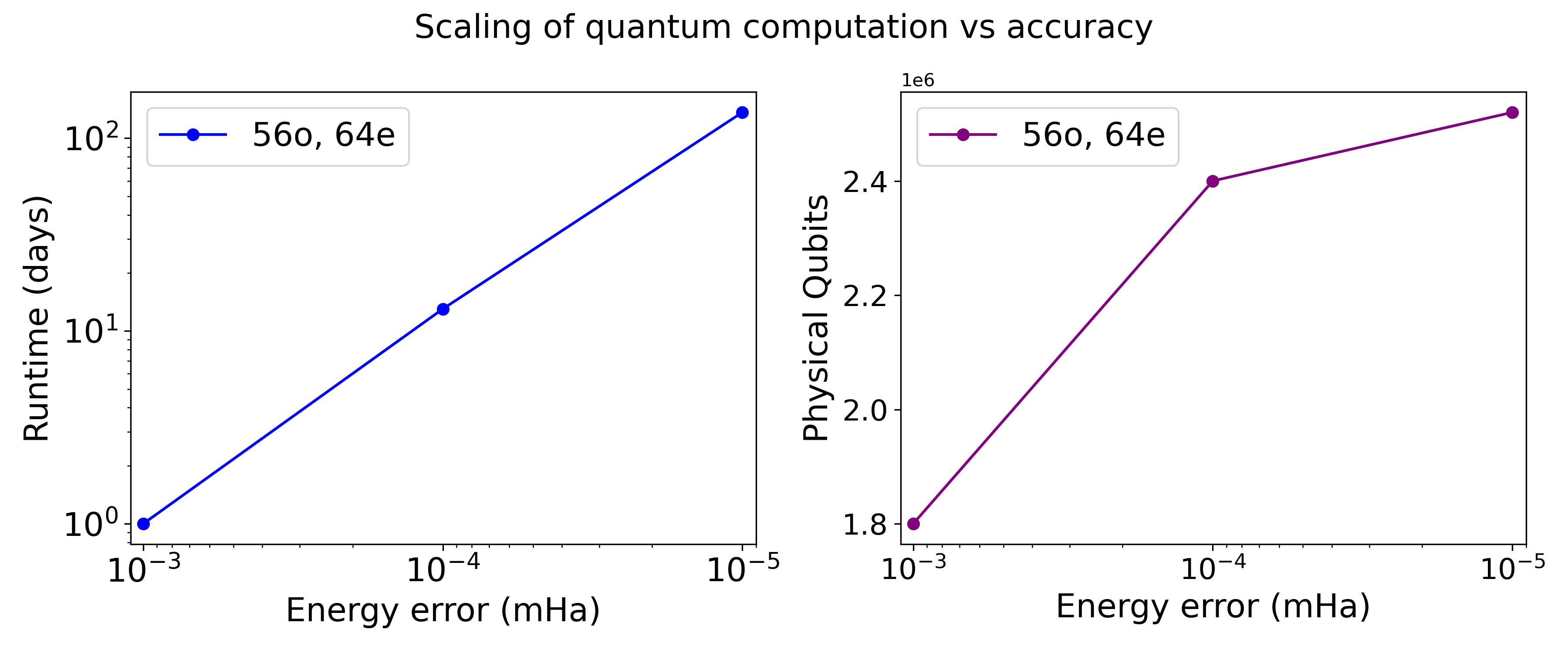}
    \caption{We report the scaling of the overall runtime and the number of physical qubits needed for the ion-trap basic hardware modeling with an increasing accuracy of the target energy error of the quantum computation for the system size of $\{56o, 64e\}$.}
    \label{Fig.18-accuracy_vs_orbitals}
\end{figure}

Here we explore the scaling of the overall runtime and the number of physical qubits needed for the basic hardware modeling versus the energy error of the calculation. This calculation incorporates all the advances shown in Fig.~\ref{Fig.8-Detailed_results} and further explores the effect of going beyond the target accuracy of 1e-3 mHa for the quantum computation. We observe a linear increase in the number of physical qubits as the desired energy error is decreased, while the overall runtime of the quantum computation increases exponentially. This exponential overhead arises from the increasing number of block-encoding repetitions required to amplify the signal in the quantum phase estimation subroutine, which scales as $\mathcal{O}(1/\epsilon)$ with respect to the target energy error $\epsilon$.

\subsection{ZX calculus and Graph State Compilation resource reductions}~\label{Subsec:resource_reductions}

Here we discuss in more detail the effect of ZX calculus with graph state compilation. Towards that end, we use a smaller QASM circuit with $129$ logical qubits, $7261$ T-gate counts per block encoding and with $3017$ number of block encodings needed. 

The rotation-aware transpiler from Bench-Q~\cite{benchq} gives a total logical T count of $8.35\times10^{7}$ for the original circuits versus $6.67\times10^{7}$ after ZX calculus incorporation~\cite{kissinger2020Pyzx}. The $18.2\%$ reduction is reasonable when applying a single full reduce pass in PyZX~\cite{kissinger2020Pyzx, Duncan2019ZX}. T-gate cancellation in PyZX’s full reduce procedure relies on optimizing the ZX-diagram representation of a quantum circuit. Since the input circuit is already written using only Clifford and T gates, PyZX can find and remove many unnecessary T gates, especially by combining or canceling them. This kind of reduction is typical and expected for large circuits~\cite{kissinger2020Pyzx, Duncan2019ZX}.

In surface-code-based architectures, T gates are not native operations and must be executed through ancilla-based protocols involving magic state injection, entanglement via CNOTs, and measurement. As a result, each T gate typically introduces multiple 2-qubit gates into the circuit. Additionally, beyond T gates, in graph-state compilation, even Clifford operations such as Hadamards and controlled rotations are implemented by preparing entangled graph structures using CZ gates and ancillary qubits.

Consequently, the T-gate counts of $7261$ per block encoding at the algorithmic level correspond to 10880 two-qubit gate count once accounted the cumulative overhead from both T gate implementation and the additional entangling gates used throughout the algorithm. The T gate count was obtained by parsing the synthesized Clifford+T circuit output directly from the pyLIQTR~\cite{Obenland_pyLIQTR} framework which compiles the quantum algorithm down to its logical gate representation before any error correction overhead is introduced. In contrast, the 2-qubit gate count was measured after compiling each phase estimation block and analyzing the full QASM circuit—including ancilla logic—prior to ZX-calculus simplification. We summarize the results in Table~\ref{Table ZX logical}

\begin{table}[h!]
    \centering
    \vspace{6pt}
    \begin{tabular}{|l|l|l|}
    \hline
    \textbf{total per block encoding} & \textbf{before ZX calculus} & \textbf{after ZX calculus} \\
    \hline
    T-gate & $2.75e04$ & $2.25e04$   \\
    \hline
    CNOT  & 10880 & 363 \\
    \hline
    remote CNOT  & 9529 & 278 \\
    \hline
    \end{tabular}
    \caption{We report logical counts on the qasm circuits before and after the implementation of ZX calculus.}
    \label{Table ZX logical} 
\end{table}

The second row of Table~\ref{Table ZX logical} shows a $30x$ reduction in the total number of two-qubit gates per block, dropping from 10,880 to 363. In this work, remote CNOTs are defined as CNOT gates between qubit pairs whose indices differ by more than 4, i.e.,  $|q_i - q_j| > 4$.  We report a $34x$ reduction in remote CNOTs (from 9529 to 278). This suggests that the percentage of remote CNOTs to the total CNOTs per block encoding before and after ZX calculus drops from $87.6 \%$ to $75.1\%$.

The significant reductions in Table~\ref{Table ZX logical} is a direct result of PyZX’s ability to apply advanced ZX-calculus-based simplifications~\cite{kissinger2020Pyzx, Duncan2019ZX}. The advantage becomes especially significant in the context of synthesized QASM~\cite{kissinger2020Pyzx, Duncan2019ZX}, which is exactly the case in our workflow. In this setting, the ZX-calculus becomes particularly powerful because it can operate directly on the logical structure of the circuit, identifying redundancies and algebraic identities that are often obscured in rigid gate sequences. This flexibility allows PyZX to optimize non-locally and leads to deep compression in both gate depth and gate count, without changing the semantics of the quantum algorithm~\cite{kissinger2020Pyzx, Duncan2019ZX}.

\begin{table}[h!]
    \centering
    \vspace{6pt}
    \begin{tabular}{|l|l|l|}
    \hline
    \textbf{} & \textbf{before ZX calculus} & \textbf{after ZX calculus} \\
    \hline
    compiled subroutines & 20 & 2 \\
    \hline
    Total cycles & $9.11e08$ & $3.83e08$ \\
    \hline
    \end{tabular}
    \caption{We report the implications of ZX calculus in GSC total number of compiled subroutines and total cycles.}
    \label{Table ZX-GSC} 
\end{table}

The ZX calculus has significant positive implications on the graph state compilation scheme (see Table~\ref{Table ZX-GSC}). In Bench-Q~\cite{benchq}, compiled subroutines refer to the distinct quantum circuit modules that make up a full algorithm and are accessed via \texttt{quantum\_program.subroutines}. Before PyZX optimization, a typical circuit may include many such subroutines (e.g., 20), reflecting higher structural complexity. After applying ZX-calculus-based simplification, this number can drop significantly (e.g., to 2), as PyZX collapses redundant logic and merges circuit operations non-locally. This reduction simplifies scheduling, improves layout allocation, and reduces the number of quantum error correction (QEC) cycles needed (see Table~\ref{Table ZX-GSC}), ultimately contributing to more compact and resource-efficient fault-tolerant execution. 

\begin{table}[h!]
    \centering
    \vspace{6pt}
    \begin{tabular}{|l|l|l|}
    \hline
    \textbf{QEC Task Allocation} & \multicolumn{2}{c|}{\textbf{Total cycles}}  \\
    \hline
    \textbf{} & \textbf{before ZX calculus} & \textbf{after ZX calculus} \\
    \hline
    T measurement & $1.97e07$ & $8.31e06$ \\
    \hline
    distillation & $8.80e08$ & $3.75e08$ \\
    \hline
    graph state prep & $6.64e04$ & $3.62e04$ \\
    \hline
    graph state prep \& distillation & $1.12e07$ & $3.26e05$ \\
    \hline
    \end{tabular}
    \caption{We report the QEC-layer surface code cycles before and after the implementation of ZX calculus.}
    \label{Table cycle reductions} 
\end{table}

For a more detailed analysis, in Table~\ref{Table cycle reductions} we report the explicit reductions in distinct QEC task allocations. By eliminating redundant T gates and compressing Clifford + T segments, the number of T measurements drops by approximately $58\%$, and the required magic state distillation cycles are reduced by nearly $57\%$. Furthermore, the simplification of entanglement patterns through ZX rewrite rules leads to fewer and smaller graph states, reducing graph state preparation cycles by about $45\%$. Lastly, tasks that involve both distillation and graph state preparation—typically costly due to their concurrency—are reduced by roughly $97\%$ thanks to restructured subroutines and fewer teleportation edges~\cite{benchq}. Overall, the reductions reflect how ZX calculus enables a more compact, hardware-friendly circuit layout, significantly lowering the resource cost across all critical fault-tolerant QEC tasks.

\subsection{Inter-ELU Entanglement in Ion-Trap Hardware}~\label{subsec:entanglement}

We explicitly model the detailed ion-trap hardware implementation at the inter-ELU level. The main objective is to estimate the number of trapped ions required to support fault-tolerant lattice surgery between surface codes, assuming a realistic time budget of $10^{-4}$ seconds per surface code cycle (SCC).

The work of Devitt et al.~\cite{Simon} highlights a significant challenge: the low probability of successful entanglement between ions in separate ELUs, quantified as $P = 2.18 \times 10^{-4}$ is fundamentally constrained by the physics of ion-photon interactions. Their model incorporates the following expression for the heralded entanglement success probability~\cite{Stephenson2020}:
\begin{equation}
P = p_{\text{Bell}} \left[ P_{\downarrow} P_e P_S \bar{P}_{\text{click}} \right]^2,
\end{equation}
where $p_{\text{Bell}} = 1/2$ reflects the fact that only two of the four Bell states are heralded; $P_{\downarrow} \approx 0.99$ is the probability of preparing the correct initial ground state; $P_e \approx 0.97$ is the excitation probability to the relevant excited state; $P_S \approx 0.95$ is the decay probability back to the $S_{1/2}$ ground state; and $\bar{P}_{\text{click}} \approx 0.023$ is the photon detection efficiency per ion. Given this baseline, the total entanglement probability per attempt is
\[
P = 2.18 \times 10^{-4},
\]
and with an excitation attempt rate of 833 kHz (including Doppler cooling overhead), this corresponds to a heralded entanglement rate of approximately 182 events per second.

Recent experimental advances report that photon collection efficiencies as high as 10\% are achievable~\cite{Kim2011PhotonCollection}. Incorporating this improved value, while holding all other parameters fixed, modifies the entanglement probability as follows:
\[
P = p_{\text{Bell}} \left[ P_{\downarrow} P_e P_S \bar{P}_{\text{click}} \right]^2 \approx \frac{1}{2} (0.99 \times 0.97 \times 0.95 \times 0.1)^2 \approx 4.16 \times 10^{-3}.
\]
This near-order-of-magnitude improvement means that with just 100 entanglement attempts (e.g., in parallelized hardware), the success probability for at least one entanglement reaches sufficiently high confidence to meet the timing requirements of $10^{-4}$ SCC durations in fault-tolerant protocols. In our analysis, we incorporate the state-of-the-art entanglement probability value of 4.16e-3.

\subsection{Classical simulations}~\label{Sec:Evidence_classical_benchmarking}

From all the methods discussed in \cite{complex_XVIII}, DFT has the best scaling of $\mathcal{O}(N^3)$ with the number of orbitals $N$, but importantly scaling with accuracy is unknown. Given  that DFT predicts the wrong sign suggests that the method fails to even find the relative ordering correct (see Table~\ref{Table XVIII classical} in Sec.~\ref{Sec:Impact}) and renders DFT for our problem unreliable. 

For the HF method, the scales even worse than DFT with $\mathcal{O}(N^4)$ and does not include electron correlation effects, while for DMRG we have $\mathcal{O}(\chi^3 d^3)$ where $\chi$ is the bond dimension and $d$ is the local site dimension. In the worst case, DMRG has exponential scaling in $\chi$, unless there is a probable area law entanglement growth. Finally, selected CI could give the best classical accuracy possible within heuristic classical methods in general, but with the price of scaling as $\mathcal{O}(N_{det} \times N_{orb}^4)$, where $N_{det}$ = number of selected determinants and $N_{orb}$ = number of orbitals. The number of determinants, $N_{det}$, generally scales exponentially with system size. In a nutshell, SHCI and DMRG scale exponentially in the worst case, but in practice can be tractable for moderate sized problems.

In general, all classical algorithms have a hidden exponential cost that the different methods try to soften. On the contrary, quantum algorithms scaling with system size is provably polynomial. In our work, the quantum algorithm used has an observed scaling of $N^{0.96}$ on the T-gate counts supported by benchmarking simulations on molecules up to N=150 orbitals~\cite{DFTHC}. This is a significant improvement to the previous best quantum algorithm with a gate complexity scaling as $N^3$. Importantly, it also has a linear scaling with the number of logical qubits. Therefore, only from complexity arguments the quantum algorithm is expected to have a more favorable scaling compared to classical. Importantly though, we fully compiled the QASM circuits to show the polynomial scaling cost with system size of the quantum algorithm under study.

Some of the key data supporting our classical benchmarking analysis are already presented in Sec.~\ref{Sec:Classical_Benchmarking} and we summarize them here for clarity.

\begin{figure}[h!]
    \centering
    \includegraphics[width=0.7\textwidth]{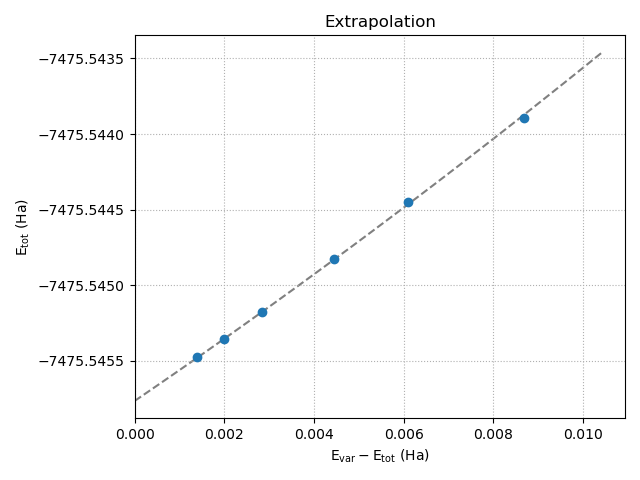}
    \caption{This plot shows the extrapolated energy to zero error of classical computation of the complex XVIII systems for $\{100o, 100e\}$ with the state-of-the-art method, SHCI. The heuristically observed uncertainty of the SHCI method for the $\{100o, 100e\}$ is $0.3mHa$, but the accuracy can not reliably be validated.}
    \label{Fig.23-100o_extrapolate}
\end{figure}

\begin{figure}[h!]
    \centering
    \includegraphics[width=0.7\textwidth]{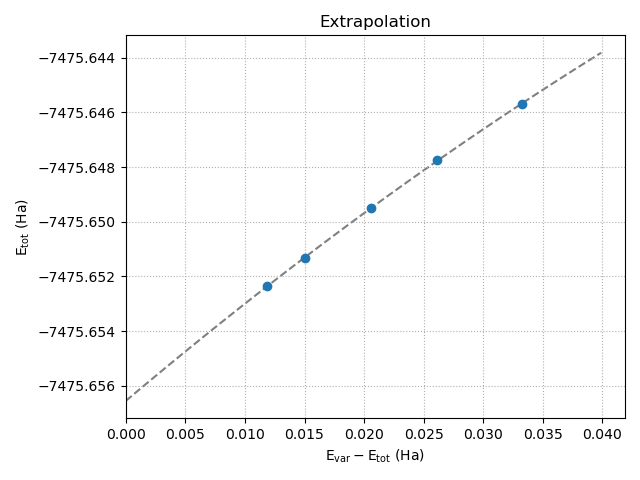}
    \caption{This plot shows the extrapolated energy to zero error of classical computation of the complex XVIII systems for $\{150o, 150e\}$ with the state-of-the-art method, SHCI. The heuristically observed uncertainty of the SHCI method for the $\{150o, 150e\}$ is $7mHa$, but the accuracy can not reliably be validated.}
    \label{Fig.24-150o_extrapolate}
\end{figure}

\begin{itemize}
    \item Fig.~\ref{Fig.6_Extrapolation} shows the SHCI extrapolated energy results for the $\{56o, 64e\}$ system with heuristic uncertainty of 0.05 mHa.
    \item Equation~\ref{CPU_equation} defines our CPU-hour calculation:
    \[
    \text{CPU-hours} = \text{Wall-clock time (hours)} \times \text{Number of CPU cores}
    \]
    \item Table~\ref{Table II} and the surrounding discussion provide the CPU-hours for larger systems ($\{100o, 100e\}$ and $\{150o, 150e\}$), estimated using consistent wall-time and core-count methodology.
    \item Systematic accuracy degradation with system size is discussed in Sec.~\ref{Sec:Evidence_classical_benchmarking} and shown in Fig.~\ref{Fig.20_classical_uncertainty}.
    \item In Table~\ref{Table Small molecules}, we report the energy estimations of the small molecules relevant for the sub-reaction shown in Eq.~\eqref{Eq:DE} within the catalytic cycle.
    \item In Table~\ref{Table DE}, we report the explicit values of the of $\Delta E^{\mathrm{el}}_{\mathrm{XVIII}}$ of Eq.~\eqref{Eq:DE} shown in Fig.~\ref{Fig.25-delta_E_XVIII.png}
    \item Fig.~\ref{Fig.23-100o_extrapolate} and Fig.~\ref{Fig.24-150o_extrapolate} shows the extrapolated energy to zero error of classical computation of the complex XVIII systems for $\{100o, 100e\}$ and $\{150o, 150e\}$ with heuristically observed uncertainties of 0.3 mHa and 7 mHa, respectively.
\end{itemize}

\begin{figure}[h!]
    \centering
    \includegraphics[width=1.0\textwidth]{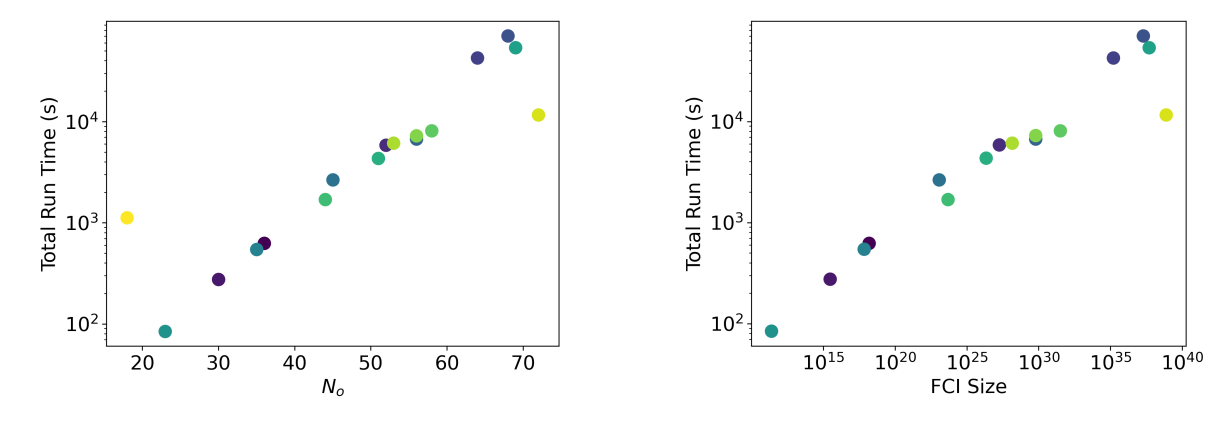}
    \caption{Total runtime (in seconds) vs number of spin orbitals \( N_o \) and FCI size for selected classical SHCI simulations.}
    \label{Fig.27-classical-runtime_vs}
\end{figure}

\begin{figure}[h!]
    \centering
    \includegraphics[width=0.6\textwidth]{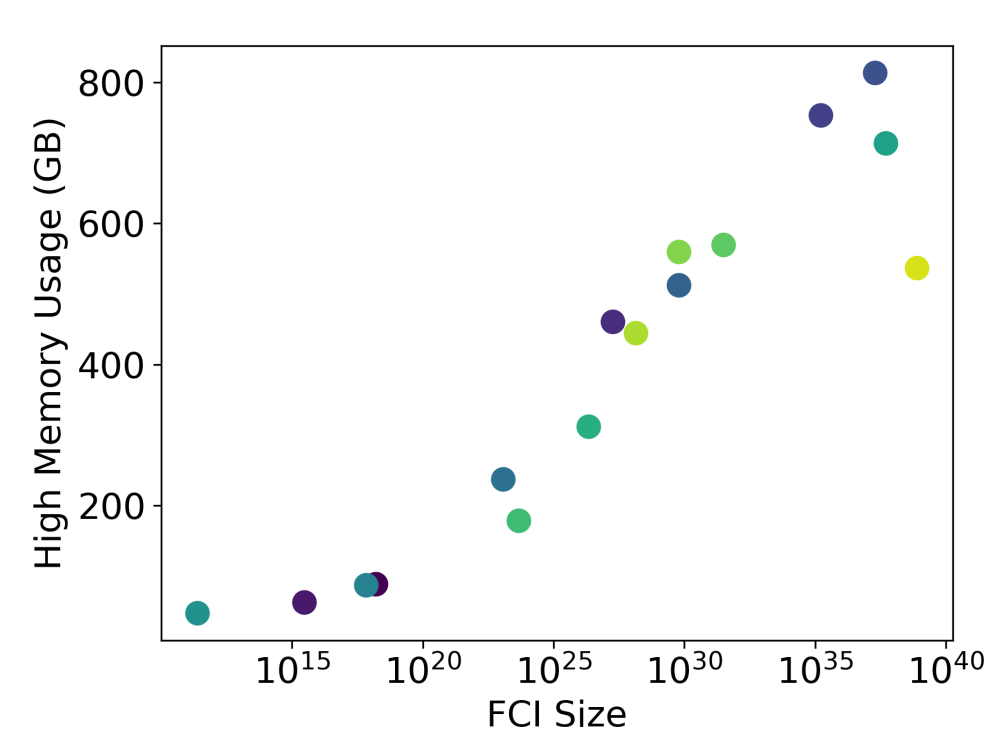}
    \caption{Total memory usage vs number of spin orbitals \( N_o \) for selected classical SHCI simulations.}
    \label{Fig.26-classical-memory-usage}
\end{figure}

Moreover, to further strengthen the arguments, we present few extra plots here:

\begin{itemize}
    \item Fig.~\ref{Fig.27-classical-runtime_vs} shows the total runtime (in seconds) of SHCI calculations as a function of the number of spin orbitals \( N_o \), demonstrating the exponential scaling of classical methods in practice. This figure is a complimentary analysis for different systems and the analysis supports the exponential scaling of the classical runtime.
    \item Fig.~\ref{Fig.26-classical-memory-usage} shows the total memory usage for different systems with an increasing FCI size. This figure stress the large  computational resources that classical methods need for large system sizes. In future work, we will perform an analysis on the memory usage of our studied systems.
\end{itemize}

These results motivate future work to further push the limits of classical computations and leverage massively parallel resources, while simultaneously highlighting where a fault-tolerant quantum solution could deliver transformative advantages.

\begin{table}[h!]
    \centering
    \vspace{6pt}
    \begin{tabular}{|l|l|}
    \hline
    \textbf{Compound} & \textbf{$E^{\text{el}}_{\text{M06--L}}$} \\
    \hline
    H$_2$      & -1.16721 \\
    \hline
    H$_2$O     & -76.35053 \\
    \hline
    CO$_2$     & -188.43534 \\
    \hline
    \end{tabular}
    \caption{M06-L/def2-SVP electronic energies for the optimized structures given in Hartree atomic units reproduced from Ref.~\cite{complex_XVIII}.}
    \label{Table Small molecules} 
\end{table}

\begin{table}[h!]
    \centering
    \vspace{6pt}
    \begin{tabular}{|l|l|}
    \hline
    \textbf{Method} & \textbf{$\Delta E^{\text{el}}_{\text{XVIII}}$} \\
    \hline
    DFT     &                               -0.02548 \\
    \hline
    HF     &                               0.420111 \\
    \hline
    CASSCF &                               0.41225 \\
    \hline
    DMRG   &                               0.441383 \\
    \hline
    SHCI (our work)  &                               0.443483 \\
    \hline
    \end{tabular}
    \caption{Calculated $\Delta E^{\text{el}}_{\text{XVIII}}$ for different methods (in Hartree) as reported in Ref.~\cite{complex_XVIII} and calculated in our work with SHCI method.}\label{Table DE} 
\end{table}

\clearpage

\bibliographystyle{unsrt}
\bibliography{main}

\end{document}